\documentstyle[preprint,aps]{revtex}
\tighten
\include{epsf}
\begin{document}
\preprint{Preprint Numbers:\parbox[t]{50mm}{ADP-96-6/T211\\
		 ANL-PHY-8415-TH-96\\ FSU-SCRI-96-38\\hep-ph/9604402}}
\draft
\title{Renormalization and Chiral Symmetry Breaking in
       Quenched QED in Arbitrary Covariant Gauge}
\author{
  Frederick T.\ Hawes\footnotemark[1],
  Anthony G.\ Williams\footnotemark[2],
	  and
  Craig D.\ Roberts\footnotemark[3]
  \vspace*{2mm} }

\address{
  \footnotemark[1] Department of Physics and SCRI, 
    Florida State University, \\ 
    Tallahassee, Florida 32306-3016
  \vspace*{2mm}\\
  \footnotemark[2] Institute for Theoretical Physics and
    Department of Physics and Mathematical Physics, \\
    University of Adelaide, 5005, Australia
  \vspace*{2mm}\\
  \footnotemark[3] Physics Division, Argonne National Laboratory,
    Argonne, Illinois 60439-4843
}
%
\maketitle
%
\begin{abstract}
  We extend a previous Landau-gauge study of subtractive renormalization
  of the fermion propagator Dyson-Schwinger equation (DSE) in strong-coupling,
  quenched QED$_4$ to {\it arbitrary} covariant gauges.  We use the
  fermion-photon proper vertex proposed by Curtis and Pennington with an
  additional correction term included to compensate for the small
  gauge-dependence induced by the ultraviolet regulator.   We discuss the
  chiral limit and the onset of dynamical chiral symmetry
  breaking in the presence of nonperturbative renormalization.  We extract
  the critical coupling in several different gauges and find evidence of a
  small residual gauge-dependence in this quantity.
\end{abstract}

\section{Introduction}
\label{sec_intro}
Strong coupling QED$_4$ has been studied for some time
within the Dyson-Schwinger equation (DSE) formalism
both for its intrinsic interest
and also as the basis for abelianized models
of nonperturbative phenomena
in technicolor theories and QCD.
For recent reviews of Dyson-Schwinger equations and their application
and numerous references see for example
Refs.~\cite{TheReview,MiranskReview,FGMS}.
The usual approach is to write the DSE for the fermion propagator or
self-energy, possibly including equations for the photon vacuum
polarization or the fermion-photon proper vertex.
In a recent study \cite{qed4_hw} it was shown for the first time how to
implement
nonperturbative renormalization in a numerical way within the DSE formalism.
In that work the calculations were carried out in quenched approximation
in Landau gauge.  Here we will extend these studies to arbitrary covariant
gauges and will also discuss the chiral limit.

The DSE's are an infinite tower of coupled integral equations and so it
is always necessary to truncate this tower at some point and
introduce an {\it Ansatz\/} for any necessary undefined Green's
functions.  It is of course important to ensure that this {\it Ansatz\/}
be consistent with all appropriate symmetries of the theory
and that it have the correct perturbative limit.
The resulting nonlinear integral equations are solved
numerically in Euclidean space by iteration.
Dynamical (or spontaneous) chiral symmetry breaking (DCSB) occurs
when the fermion propagator develops a nonzero scalar self-energy
in the absence of an explicit chiral symmetry breaking (ECSB)
fermion mass.
We will refer to coupling constants strong enough to induce DCSB
as supercritical and those weaker are called subcritical.
We write the fermion propagator as
\begin{equation}                        \label{fermprop_formal}
  S(p) = \frac{Z(p^2)}{\not\!p - M(p^2)}
       = \frac{1}{A(p^2) \not\!p - B(p^2)}
\end{equation}
where we refer to $A(p^2)\equiv 1/Z(p^2)$
as the finite momentum-dependent fermion renormalization
and where $M(p^2)\equiv B(p^2)/A(p^2)$ is the fermion mass function.
In the massless theory
(i.e., in the absence of an ECSB bare fermion mass)
by definition DCSB occurs when $M(p^2)\neq 0$. 

Until relatively recently, most studies have used the
bare vertex as an {\it Ansatz\/} for the one-particle
irreducible (1-PI) vertex $\Gamma^\nu(k,p)$,
\cite{Bare,Miransk1,Miransk2,Miransk3,Rakow}
despite the fact that this violates the Ward-Takahashi Identity (WTI)
\cite{WTI}.  The resulting fermion propagator
is not gauge-covariant, i.e., physical quantities such as the
critical coupling for dynamical symmetry breaking and the fermion
mass pole are gauge-dependent \cite{ABKHN,CPIV,KKM}.
There have been several studies which attempted
to make the fermion DSE gauge-covariant
by using improved vertex forms which satisfy the WTI
but which possess kinematic singularities in the limit of zero
photon momentum \cite{KKM,bad_Gamma}.
A general form for $\Gamma^\nu(k,p)$ which does satisfy the Ward
Identity and which has no unphysical kinematic singularities
was given by Ball and Chiu in 1980 \cite{BC};  it consists of
a minimal longitudinally constrained term which satisfies the WTI,
and a set of tensors spanning the subspace
transverse to the photon momentum $q$.

While the WTI is necessary for gauge-invariance, it is not a
sufficient condition and in itself does not ensure gauge covariance
of the fermion propagator.  Furthermore, with many vertex
{\it Ans\"{a}tze\/} the fermion propagator DSE is not multiplicatively
renormalizable, which is equivalent to saying that
overlapping logarithms are present.
There has been much recent research on the use
of the transverse parts of the vertex to ensure both gauge-covariant 
and multiplicatively renormalizable solutions
\cite{CPIV,King,gaugetech,HaeriQED,CPI,CPII,CPIII,dongroberts,BP1,BP2},
some of which will be discussed below.

With the exception of Ref.~\cite{qed4_hw}, studies have mostly neglected
the issue of the subtractive renormalization of the DSE for the fermion
propagator. Typically these studies have assumed an initially massless theory
and have renormalized at the ultraviolet cutoff of the loop integration,
taking $Z_1 = Z_2 = 1$.  Where a nonzero bare mass has been used,
it has simply been added to the scalar term in the
propagator.  While in some circumstances for the special case of
Landau gauge this can be a reasonable approximation, it is in
general incorrect.  Although there have been earlier formal
discussions of renormalization \cite{MiranskReview,CPIV,CPI},
the important step of subtractive renormalization had not been performed
prior to the recent study in Landau gauge \cite{qed4_hw}.

Here we present the results of a study of subtractive
renormalization in the fermion DSE in {\it arbitrary} covariant gauge
for quenched strong-coupling QED$_4$.
Note that here the term ``quenched'' means
that the bare photon propagator is used in the fermion self-energy DSE,
so that $Z_3 = 1$ and there is no renormalization of the electron charge.
This is a somewhat different usage to that found in lattice gauge
theory studies, since in our study virtual fermion loops may still
be present in the proper fermion-photon vertex.

The organization of the paper is as follows:
The formalism is discussed in Sec.~\ref{sec_formalism}.
This section contains discussions of the DSE for the renormalized
fermion propagator, the {\it Ans\"{a}tze\/} for the proper vertex,
the subtractive renormalization procedure, the chiral limit, and
renormalization point transformations.
Our detailed numerical results are presented in Sec.~\ref{sec_results} 
and we present our summary and conclusions in Sec.~\ref{sec_conclusions}.

\section{Formalism}
\label{sec_formalism}
\subsection{Renormalized DSE}
\label{subsec_ren_DSE}
The DSE for the renormalized fermion propagator, in an arbitrary covariant
gauge, is
\begin{equation} \label{fermDSE_eq}
  S^{-1}(p) = Z_2(\mu,\Lambda)[\not\!p - m_0(\Lambda)]
    - i Z_1(\mu,\Lambda) e^2 \int^{\Lambda} \frac{d^4k}{(2\pi)^4}
	  \gamma^{\mu} S(k) \Gamma^{\nu}(k,p) D_{\mu \nu}(q)\:;
\end{equation}
here $q=k-p$ is the photon momentum, $\mu$ is the renormalization
point, and $\Lambda$ is a regularizing parameter (taken here to be an
ultraviolet momentum cutoff).  We write
$m_0(\Lambda)$ for the regularization-parameter dependent bare mass.
The renormalized charge is $e$ (as opposed to the bare charge $e_0$),
and the general form for the renormalized photon propagator is
\begin{equation}
  D^{\mu\nu}(q) = \left\{
    \left( -g^{\mu\nu} + \frac{q^\mu q^\nu}{q^2} \right)
    \frac{1}{1+\Pi(q^2)} - \xi \frac{q^\mu q^\nu}{q^2} \right\}
    \frac{1}{q^2}\:,
\end{equation}
with $\xi$ the renormalized covariant gauge parameter
and $\xi_0 \equiv Z_3(\mu,\Lambda)\xi$ the corresponding bare one.
Since we work in the quenched approximation, we have
for the coupling strength and gauge parameter respectively
$\alpha\equiv e^2/4\pi = \alpha_0\equiv e_0^2/4\pi$ and $\xi=\xi_0$,
and for the photon propagator we have
\begin{equation}
  D^{\mu\nu}(q) \to  D_0^{\mu\nu}(q) 
    =  \left\{\left( -g^{\mu\nu} + \frac{q^\mu q^\nu}{q^2} \right)
          - \xi \frac{q^\mu q^\nu}{q^2} \right\}\frac{1}{q^2}\:.
\end{equation}

\subsection{Vertex Ansatz}
\label{subsec_QEDvtx}

The requirement of gauge invariance in QED leads to a set of identities
referred to as the Ward-Takahashi
Identities (WTI).  The WTI for the fermion-photon vertex is
\begin{equation}
  q_\mu \Gamma^\mu(k,p) = S^{-1}(k) - S^{-1}(p)\;,
\label{WTI}
\end{equation}
where $q = k - p$\/ .  This is a generalization of the
original differential Ward identity, which expresses the effect of
inserting a zero-momentum photon vertex into the fermion propagator,
\begin{equation}
  \frac{\partial S^{-1}(p)}{\partial p_\nu} = \Gamma^{\nu}(p,p)\:.
\label{WI}
\end{equation}
The Ward identity Eq.~(\ref{WI}) follows immediately from the WTI of 
Eq.~(\ref{WTI}) after setting to zero all but the $\nu$ component of $q$,
dividing both sides of the WTI by $q_\nu$ and then taking $q_\nu\to 0$. 
In general, for nonvanishing photon momentum $q$, only the longitudinal
component of the proper vertex is constrained, i.e., the WTI provides
no information on $\Gamma^\nu_{\rm T}(k,p)\equiv
{\cal T}^{\mu\nu}\Gamma_\nu(p,k)$ for $q\neq 0$.  [We use the notation
${\cal T}^{\mu\nu}\equiv g^{\mu\nu}-(q^\mu q^\nu/q^2)$ and
${\cal L}^{\mu\nu}\equiv (q^\mu q^\nu/q^2)$ for the transverse and
longitudinal projectors respectively.]
In particular, the WTI guarantees the equality of the propagator
and vertex renormalization constants, $Z_2 \equiv Z_1$ (at least in any
reasonable subtraction scheme \cite{TheReview}.)
The WTI can be shown to be satisfied
order-by-order in perturbation theory and can also be derived
nonperturbatively.

As discussed in \cite{TheReview,Craig_1}, this can be thought
of as just one of a set of six general requirements on the vertex:
(i) the vertex must satisfy the WTI; (ii) it should contain no kinematic
singularities; (iii) it should transform under charge conjugation ($C$),
parity inversion ($P$), and time reversal ($T$) in the same way
as the bare vertex, e.g.,
	\begin{equation}
	  C^{-1} \Gamma_\mu(k,p) C = - \Gamma_\mu^{\rm T}(-p,-k)
	\end{equation}
(where the superscript {\rm T} indicates the transpose);
(iv) it should reduce to the bare vertex in the weak-coupling
limit; (v) it should ensure multiplicative renormalizability of the
DSE in Eq. (\ref{fermDSE_eq});
(vi) the transverse part of the vertex should be specified to 
	ensure gauge-covariance of the DSE.

Ball and Chiu \cite{BC} have given a description of the most general
fermion-photon vertex that satisfies the WTI; it consists of a
longitudinally-constrained (i.e., ``Ball-Chiu'') part
$\Gamma^\mu_{\rm BC}$, which is a minimal solution of the WTI,
and a basis set of eight transverse vectors $T_i^\mu(k,p)$,
which span the hyperplane specified by ${\cal L}_{\mu\nu}T_i^\nu(k,p) = 0$
(i.e., $q_\nu T_i^\nu(k,p) = 0$), where $q \equiv k-p$.
The minimal longitudinally constrained part of the vertex will be referred
to as the Ball-Chiu vertex and is given by
\begin{equation} \label{minBCvert_eqn}
  \Gamma^\mu_{\rm BC}(k,p) = \frac{1}{2}[A(k^2) +A(p^2)] \gamma^\mu
    + \frac{(k+p)^\mu}{k^2-p^2}
      \left\{ [A(k^2) - A(p^2)] \frac{{\not\!k}+ {\not\!p}}{2}
	      - [B(k^2) - B(p^2)] \right\}\:.
\end{equation}
Note that since neither ${\cal L}_{\mu\nu}\Gamma^\nu_{\rm BC}(k,p)$
nor ${\cal T}_{\mu\nu}\Gamma^\nu_{\rm BC}(k,p)$ vanish identically,
the Ball-Chiu vertex has both longitudinal and transverse components.
The transverse tensors can be conveniently written as \cite{Kiz_et_al}
\begin{eqnarray}
  T_1^\mu(k,p) & = & p^{\mu}(k\cdot q)-k^{\mu}(p\cdot q)\:, 
                                                            \label{T1mu}\\
  T_2^\mu(k,p) & = & \left[p^{\mu}(k\cdot q)-k^{\mu}(p\cdot q)\right]
                     ({\not\! k}+{\not\! p})\:,             \label{T2mu}\\
  T_3^\mu(k,p) & = & q^2\gamma^{\mu}-q^{\mu}{\not \! q}\:,  \label{T3mu}\\
  T_4^\mu(k,p) & = & q^2\left[\gamma^{\mu}(\not\!p+\not\!k)-p^{\mu}-k^{\mu}
                     \right] - 2i(p-k)^{\mu}k^{\lambda}p^{\nu}
                     \sigma_{\lambda\nu}\:,                 \label{T4mu}\\
  T_5^\mu(k,p) & = & -iq_{\nu}{\sigma^{\nu\mu}}\:,          \label{T5mu}\\
  T_6^\mu(k,p) & = & \gamma^{\mu}(p^2-k^2)+(p+k)^{\mu}{\not \! q}\:,
                                                            \label{T6mu}\\
  T_7^\mu(k,p) & = & \frac{1}{2}(p^2-k^2)\left[\gamma^{\mu}({\not \! p}+
                     {\not \! k})-p^{\mu}-k^{\mu}\right]
                     - i\left(k+p\right)^{\mu}k^{\lambda}p^{\nu}
                       \sigma_{\lambda\nu}\:,               \label{T7mu}\\
  T_8^\mu(k,p) & = & i \gamma^{\mu}k^{\nu}p^{\lambda}{\sigma_{\nu\lambda}}
                     +k^{\mu}{\not \! p}-p^{\mu}{\not \! k}\:, 
                                                            \label{T8mu}
\end{eqnarray}
where we use the conventions $g^{\mu\nu}={\rm diag}(1,-1,-1,-1)$,
$\{\gamma^\mu,\gamma^\nu\}=2g^{\mu\nu}$, and
$\sigma^{\mu\nu}\equiv (i/2)[\gamma^{\mu},\gamma^{\nu}]$.
Note that these tensors have been written in a different
linear combination to the ones presented in Ref.~\cite{qed4_hw}.
A general vertex is then written as
\begin{equation} \label{anyfullG_eqn}
  \Gamma^\mu(k,p) = \Gamma_{BC}^\mu(k,p)
    + \sum_{i=1}^{8} \tau_i(k^2,p^2,q^2) T_i^\mu(k,p)\:,
\end{equation}
where the $\tau_i$ are functions that must be chosen to give the
correct $C$, $P$, and $T$ invariance properties.

Curtis and Pennington published a series of articles
\cite{CPIV,CPI,CPII,CPIII} 
describing their specification of
a particular transverse vertex term,
in an attempt to produce
gauge-covariant and multiplicatively
renormalizable solutions to the DSE.
In the framework of massless QED$_4$, they eliminated
the four transverse vectors which are Dirac-even and must
generate a scalar term.  By requiring that the vertex $\Gamma^\mu(k,p)$
reduce to the leading log result for $k \gg p$ they were led to
eliminate all the transverse basis vectors except $T_6^\mu$, with a
dynamic coefficient chosen to make the DSE multiplicatively
renormalizable.  This coefficient had the form
\begin{equation}
  \tau_6(k^2,p^2,q^2) = -\frac{1}{2}[A(k^2) - A(p^2)] / d(k,p)\:,
\label{CPgamma1}
\end{equation}
where $d(k,p)$ is a symmetric, singularity-free function of $k$ and $p$,
with the limiting behavior $\lim_{k^2 \gg p^2} d(k,p) = k^2$.
[Here, $A(p^2)\equiv 1/Z(p^2)$ is their $1/{\cal F}(p^2)$.]
For purely massless QED, they found a suitable form,
$d(k,p) = (k^2 - p^2)^2/(k^2+p^2)$.  This was generalized to the
case with a dynamical mass $M(p^2)$, to give
\begin{equation}
  d(k,p) = \frac{(k^2 - p^2)^2 + [M^2(k^2) + M^2(p^2)]^2}{k^2+p^2}\:.
  \label{CPgamma2}
\end{equation}
They then showed that multiplicative renormalizability is retained
up to next-to-leading-log order in the DCSB case.
Subsequent papers established the form of the solutions
for the renormalization and the mass \cite{CPIII}
and studied the gauge-dependence of the solutions \cite{CPIV}.
Dong, Munczek and Roberts \cite{dongroberts} subsequently showed
that the lack of exact gauge-covariance of the solutions was due
to the use of a momentum cutoff in the integral equations,
since this type of regularization is not Poincar\'{e} invariant.
The fact that there is still some residual gauge-dependence
in the physical observables such as the chiral critical point
shows that with a momentum cutoff
the C-P vertex {\it Ansatz\/} is not yet the ideal choice.
Dong, Munczek and Roberts \cite{dongroberts}
derived an {\it Ansatz\/} for the transverse vertex terms
which satisfies the WTI
and makes the fermion propagator gauge-covariant
under hard momentum-cutoff regularization.

Bashir and Pennington \cite{BP1,BP2} subsequently described
two different vertex {\it Ans\"{a}tze\/} which make
the fermion self-energy exactly gauge-covariant,
in the sense that the critical point for the chiral phase transition
is independent of gauge.
Specific constraints they have assumed for the vertex are,
in \cite{BP1} that the transverse vertex parts
vanish in the Landau gauge,
and in \cite{BP2} that the anomalous dimension of the
fermion mass function $\gamma_m$ is exactly 1 at the critical coupling.
Their work is a continuation of that of Dong, Munczek and Roberts,
and indeed their vertex {\it Ansatz\/} corresponds
to the general form suggested in \cite{dongroberts}.

However, the kinematic factors $\tau_{2,3,6,8}$ in both vertex forms
are rather complicated and depend upon a pair of as yet
undetermined functions $W_{1,2}(k^2,p^2)$ which must be chosen to
guarantee that the weak-coupling limit of $\Gamma^\mu$ matches the
perturbative result.  Renormalization studies of the DSE using these new
vertex {\it Ans\"atze\/} should be interesting and represent a direction
for further research.

For the solutions to the fermion DSE using the C-P vertex, the critical
point for the chiral phase transition has been shown to have a much
weaker gauge-dependence than that for the DSE with the bare or minimal
Ball-Chiu vertices \cite{ABGPR}. 
In this work we will use the Curtis-Pennington
{\it Ansatz\/} as the basis for our calculations.

The equations are separated into
a Dirac-odd part describing the finite propagator renormalization
$A(p^2)$, and a Dirac-even part for the scalar self-energy, by taking
$\frac{1}{4}{\rm Tr}$ of the DSE multiplied by $\not\!p/p^2$ and 1,
respectively.  The equations are solved in Euclidean space and so
the volume integrals
$\int d^4k$ can be separated into angle integrals and an integral
$\int dk^2$; the angle integrals are easy to perform analytically,
yielding the two equations which will be solved numerically.

One refinement of our treatment of the C-P vertex in the present work is
associated with subtleties in the ultraviolet regularization scheme.
Although there have been some exploratory studies of
dimensional regularization for the DSE \cite{dim_reg},
this has not yet proven practical in nonperturbative field theory
and momentum cutoffs for now remain the regularization scheme
of choice in such studies.
Naive imposition of a momentum cutoff destroys the gauge covariance
of the DSE because the self-energy integral contains terms,
related to the vertex WTI, which should vanish
but which are nonzero when integrated under cutoff regularization
\cite{dongroberts,BP1}.
In Appendix  \ref{appdx_a} we derive an expression for one such undesirable
term and show how it may subtracted in a simple way
from the regularized self-energy.
We have also calculated some DSE solutions with the usual uncorrected
UV cut-off method for comparison purposes,
but otherwise we use this ``gauge-improved'' regularization
combined with the C-P vertex throughout this work.  This will be commented
on futher in the discussion of numerical results in
Sec.~\ref{sec_results}. 

\subsection{Subtractive Renormalization}
\label{subsec_subtr}
The subtractive renormalization of the fermion propagator DSE proceeds
similarly to the one-loop renormalization of the propagator in QED.
(This is discussed in \cite{TheReview} and in \cite{IZ}, p.~425ff.)
One first determines a finite, {\it regularized\/} self-energy,
which depends on both a regularization parameter and the
renormalization point;
then one performs a subtraction at the renormalization point,
in order to define the renormalization parameters $Z_1$, $Z_2$, $Z_3$
which give the full (renormalized) theory in terms of the regularized
calculation.

A review of the literature of DSEs in QED shows, however, that this
step is never actually performed.  Curtis and Pennington \cite{CPIV}
for example, define their renormalization point at the UV cutoff.

Many studies take $Z_1 = Z_2 = 1$ \cite{CPIV,King,HaeriQED,CPI,CPII,CPIII};
this is a reasonable approximation in Landau gauge in cases where the coupling
$\alpha$ is sufficiently small (i.e.,
$\alpha$ \raisebox{-.4ex}{$\stackrel{<}{\sim}$} 1),
but if $\alpha$ is chosen large enough,
the value of the dynamical mass at the renormalization point may be
significantly large compared with its maximum in the infrared.
For instance, in Ref.~\cite{CPIV}, figures for the fermion mass are given
with $\alpha$ = 0.97, 1.00, 1.15 and 2.00 in various gauges.  For
$\alpha = 2.00$, the fermion mass at the cutoff is down by only an
order of magnitude from its limiting value in the infrared. 
In general for strong coupling and/or gauges other than Landau gauge
this approximation is unreliable.

As shown in Ref.~\cite{qed4_hw} subtractive renormalization can be
properly implemented in numerical DSE studies without such
approximations.  We begin with a summary of the renormalization
procedure \cite{TheReview,qed4_hw}.  One defines
a regularized self-energy $\Sigma'(\mu,\Lambda; p)$, leading to the
DSE for the renormalized fermion propagator,
\begin{eqnarray} \label{ren_inv_S}
  \widetilde{S}^{-1}(p) & = & Z_2(\mu,\Lambda) [\not\!p - m_0(\Lambda)]
    - \Sigma'(\mu,\Lambda; p) \nonumber\\
    & = & \not\!p - m(\mu) - \widetilde{\Sigma}(\mu;p)
      = A(p^2)\not\!p -B(p^2)\:,
\end{eqnarray}
where the (regularized) self-energy is
\begin{equation} \label{reg_Sigma}
  \Sigma'(\mu,\Lambda; p) = i Z_1(\mu,\Lambda) e^2 \int^{\Lambda}
    \frac{d^4k}{(2\pi)^4} \gamma^\lambda \widetilde{S}(\mu;k)
      \widetilde{\Gamma}^\nu(\mu; k,p)
      \widetilde{D}_{\lambda \nu}(\mu; (p-k))\:.
\end{equation}
[To avoid confusion we will follow Ref.~\cite{TheReview} and in this
section {\it only} we will denote
regularized quantities with a prime
and renormalized ones with a tilde, e.g. $\Sigma'(\mu,\Lambda; p)$
is the regularized self-energy depending on both the renormalization
point $\mu$ and regularization parameter $\Lambda$
and $\widetilde{\Sigma}(\mu;p)$ is the renormalized self-energy.]
As suggested by the notation (i.e., the omission of the $\Lambda$-dependence)
renormalized quantities must become independent of the regularization-parameter
as the regularization is removed (i.e., as $\Lambda\to\infty$).
The self-energies are decomposed into Dirac and scalar parts,
\begin{equation}
  \Sigma'(\mu,\Lambda; p) = \Sigma'_d(\mu,\Lambda; p^2) \not\!p
		     + \Sigma'_s(\mu,\Lambda; p^2)
  \label{decompose}
\end{equation}
(and similarly for the renormalized quantity,
$\widetilde{\Sigma}(\mu,p)$).
By imposing the renormalization boundary condition,
\begin{equation}
  \left. \widetilde{S}^{-1}(p) \right|_{p^2 = \mu^2}
  = \not\!p - m(\mu)\:,
\label{ren_point_BC}
\end{equation}
one gets the relations
\begin{equation}\label{ren_BC}
  \widetilde{\Sigma}_{d,s}(\mu; p^2) =
    \Sigma'_{d,s}(\mu,\Lambda; p^2) - \Sigma'_{d,s}(\mu,\Lambda; \mu^2) 
\end{equation}
for the self-energy,
\begin{equation}
  Z_2(\mu,\Lambda) = 1 + \Sigma'_d(\mu,\Lambda; \mu^2)
\label{eq_Z2}
\end{equation}
for the renormalization constant, and
\begin{equation}
  m_0(\Lambda) = \left[ m(\mu) - \Sigma'_s(\mu,\Lambda; \mu^2) \right]
	/ Z_2(\mu,\Lambda)
\label{baremass}
\end{equation}
for the bare mass.
The mass renormalization constant is then given by
\begin{equation}
  Z_m(\mu,\Lambda) = m_0(\Lambda)/m(\mu)\:,
\label{Z_m}
\end{equation}
i.e., as the ratio of the bare to renormalized mass.

The vertex renormalization, $Z_1(\mu,\Lambda)$, is identical to
$Z_2(\mu,\Lambda)$ as long as the vertex {\it Ansatz\/} satisfies
the Ward Identity; this is how it is recovered for multiplication
into $\Sigma'(\mu,\Lambda;p)$ in Eq. (\ref{reg_Sigma}).

In order to obtain numerical solutions, the final Minkowski-space
integral equations are first rotated to Euclidean space.  [Note that
all equations in Secs.~\ref{sec_intro} and \ref{sec_formalism}
are written in Minkowski space.]  They are
then solved by iteration on a logarithmic grid from an initial guess.
The solutions are confirmed to be independent of the initial guess
and are solved with a wide range of cutoffs ($\Lambda$), renormalization
points ($\mu$), couplings ($\alpha$), covariant gauge choices ($\xi$),
and renormalized masses ($m(\mu)$).

%
The chiral limit occurs by definition
when the bare mass is taken to zero sufficiently rapidly as
the regularization is removed.  This is guaranteed, for example, by
maintaining $m_0(\Lambda) = 0$ as $\Lambda \to \infty$.
Explicit chiral symmetry breaking (ECSB) occurs
when the bare mass $m_0(\Lambda)$ is not zero, (or more precisely,
whenever it is not taken to zero sufficiently rapidly as $\Lambda \to
\infty$).  Dynamical chiral symmetry breaking (DCSB) is said to have occured
when $M(p^2) \not = 0$ in the absence of ECSB.
As the coupling strength increases from zero,
there is a transition to a DCSB phase
at the critical coupling $\alpha_c$.
Concisely, the absence of ECSB means that we can set $m_0(\Lambda) = 0$,
and the absence of both ECSB and DCSB (i.e., $\alpha < \alpha_c$)
means that $M(p^2)$, $m(\mu)$, and $m_0(\Lambda)$ simultaneously vanish.
(Recall that in the notation that we use here,
$M(p^2) \equiv B(p^2)/A(p^2)$ and $m(\mu) \equiv M(\mu^2)$.)
This is the same definition of the chiral limit that is used in
nonperturbative studies of QCD,
see e.g.\ Refs.~\cite{TheReview,MiranskReview,FGMS,Rothe}
and references therein.
Obviously, any limiting procedure where we take $m_0(\Lambda) \to 0$
sufficiently rapidly as $\Lambda \to \infty$
will also lead to the chiral limit \cite{MiranskReview}.

\subsection{Renormalization Point Transformations}
\label{subsec_ren_pt}

A renormalization point transformation is a change of renormalization scale
[i.e., ${\cal R}(\mu,\mu')$ for $\mu\to\mu'$] such that the bare mass(es) and
coupling(s) remain fixed for fixed regularization parameter ($\Lambda$) and
fixed renormalization scheme.  This ensures that the physical observables of
the theory are invariant under such a transformation.  This
set of transformations is associative
[${\cal R}(\mu,\mu'){\cal R}(\mu',\mu'')={\cal R}(\mu,\mu'')$],
contains the identity [${\cal R}(\mu,\mu)={\cal I}$], and contains all
inverses [${\cal R}(\mu,\mu')^{-1}={\cal R}(\mu',\mu)$] and hence is called
the renormalization group.

For the purposes of the discussion here we will now explicitly indicate
the choice of renormalization point by a $\mu$-dependence of the renormalized
quantities, i.e., $A(\mu;p^2)\equiv 1/Z(\mu;p^2)$,
$M(\mu;p^2)\equiv B(\mu;p^2)/A(\mu;p^2)$, etc.
Note that Eq.~(\ref{ren_inv_S}) implies that 
\begin{eqnarray}
A(\mu;p^2)&=&Z_2(\mu,\Lambda)-\Sigma'_d(\mu,\Lambda;p^2)
=1-\widetilde\Sigma_d(\mu,\Lambda;p^2) \;, \nonumber\\
B(\mu;p^2)&=&Z_2(\mu,\Lambda)m_0(\Lambda)+\Sigma'_s(\mu,\Lambda;p^2)
= m(\mu)+\widetilde\Sigma_s(\mu,\Lambda;p^2) \;.
\label{AB_at_mu}
\end{eqnarray} 
The renormalization point boundary condition in
Eq.~(\ref{ren_point_BC}) then leads to
$\widetilde\Sigma(\mu,\Lambda;\mu^2)=0$,
or equivalently, to the two boundary conditions
$A(\mu;\mu^2)=1$ and $M(\mu;\mu^2)=B(\mu;\mu^2)=m(\mu)$.
From Eq.~(\ref{AB_at_mu}) and the fact that
$Z_1(\mu,\Lambda)=Z_2(\mu,\Lambda)$ we have 
\begin{eqnarray}
\left[A(\mu;p^2)/Z_2(\mu,\Lambda)\right]&=&
1-\left[\Sigma'_d(\mu,\Lambda;p^2)/Z_1(\mu,\Lambda)\right] \;,
\nonumber\\
\left[B(\mu;p^2)/Z_2(\mu,\Lambda)\right]&=&m_0(\Lambda)+
    \left[\Sigma'_s(\mu,\Lambda;p^2)/Z_1(\mu,\Lambda)\right]\;.
\label{AB_rescaled}
\end{eqnarray} 
Consider the effects of an an arbitrary rescaling $A(p^2)\to cA(p^2)$ and
$B(p^2)\to cB(p^2)$, [i.e., $M(p^2)$ fixed],  for some real constant $c$.
It is straightforward to see that under such a
rescaling we have $S(p)\to (1/c)S(p)$ and $\Gamma^\nu(p',p)\to
c\Gamma^\nu(p',p)$.
It follows that the RH sides of Eqs.~(\ref{AB_rescaled})
are unaffected by such an arbitrary rescaling.  Hence, it follows that
the choice of renormalization point boundary conditions is equivalent
to the choice of scale for the functions $A$ and $B$.  

Let us consider this observation in more detail.
Since we are working in the quenched approximation, where $e^2$ and
$\widetilde D$ are unaffected by a change of renormalization point,
it follows from Eq.~(\ref{reg_Sigma}) that
$\Sigma'(\mu,\Lambda;p^2)/Z_1(\mu,\Lambda)$
is renormalization point independent since a change of
renormalization point is a rescaling of $A$ and $B$.
Then since
$m_0(\Lambda)$ is renormalization point independent by definition,
the entire RHS of Eqs.~(\ref{AB_rescaled}) must be independent
of the choice of renormalization point.
Thus, under a renormalization point transformation we must have
{\em for all} $p^2$
\begin{eqnarray}
  M(\mu';p^2)&=&M(\mu;p^2)\equiv M(p^2) \;, \nonumber\\
  \frac{A(\mu';p^2)}{A(\mu;p^2)}&=&
     \frac{Z_2(\mu',\Lambda)}{Z_2(\mu,\Lambda)}
     =A(\mu';\mu^2)=\frac{1}{A(\mu;\mu'^2)} \;,
\label{ren_pt_transf}
\end{eqnarray}
from which it follows for the fermion propagator that 
$\widetilde S(\mu';p)/\tilde S(\mu;p)
  = Z_2(\mu,\Lambda)/Z_2(\mu',\Lambda)$
in the usual way.  The behavior in Eq.~(\ref{ren_pt_transf}) is
explicitly tested for our numerical solutions.  It is clear from
Eq.~(\ref{ren_pt_transf}) that having a solution at one renormalization
point ($\mu$) completely determines the solution at any other renormalization
point ($\mu'$) without the need for any further computation.

An alternative derivation of this result which starts
from the renormalized action and which applies to the general unquenched
case can be found for example in Sec.~2.1 of Ref.~\cite{TheReview}.
For brevity we can denote the above renormalization point dependence of
the fermion propagator by
$\widetilde S(\mu;p)\propto 1/Z_2(\mu,\Lambda)$.
In the general unquenched case \cite{IZ} we would have in addition 
$\widetilde D^{\sigma\nu}(\mu;q)
    \propto \xi(\mu)\propto 1/Z_3(\mu,\Lambda)$,
$e(\mu) \propto Z_2(\mu,\Lambda)
    \sqrt{Z_3(\mu,\Lambda)}/Z_1(\mu,\Lambda)$, and
$e(\mu)\widetilde\Gamma^\nu(\mu;q,p)
    \propto Z_2(\mu,\Lambda)\sqrt{Z_3(\mu,\Lambda)}$.

\section{Results}
\label{sec_results}

Solutions were obtained in Euclidean space for the DSE for couplings
$\alpha$ from 0.1 to 1.30, in gauges with $\xi$ from -0.25 to 3,
and with a variety of renormalization points and renormalized masses.
All results in this section refer to Euclidean space quantities.
In the graphs and tables that follow, there are no explicit mass units.
Since the equations have no inherent mass-scale, the cutoff
$\Lambda$, renormalization point $\mu$, $m(\mu)$, and units of
$M(p^2)$ or $B(p^2)$ all scale multiplicatively, and the units
are arbitrary.  In four dimensions the coupling has no mass dimension,
therefore it remains unchanged for all such choices of mass units.

\begin{figure}[htb]
  \setlength{\epsfxsize}{12.0cm}
  \centering
    \epsffile{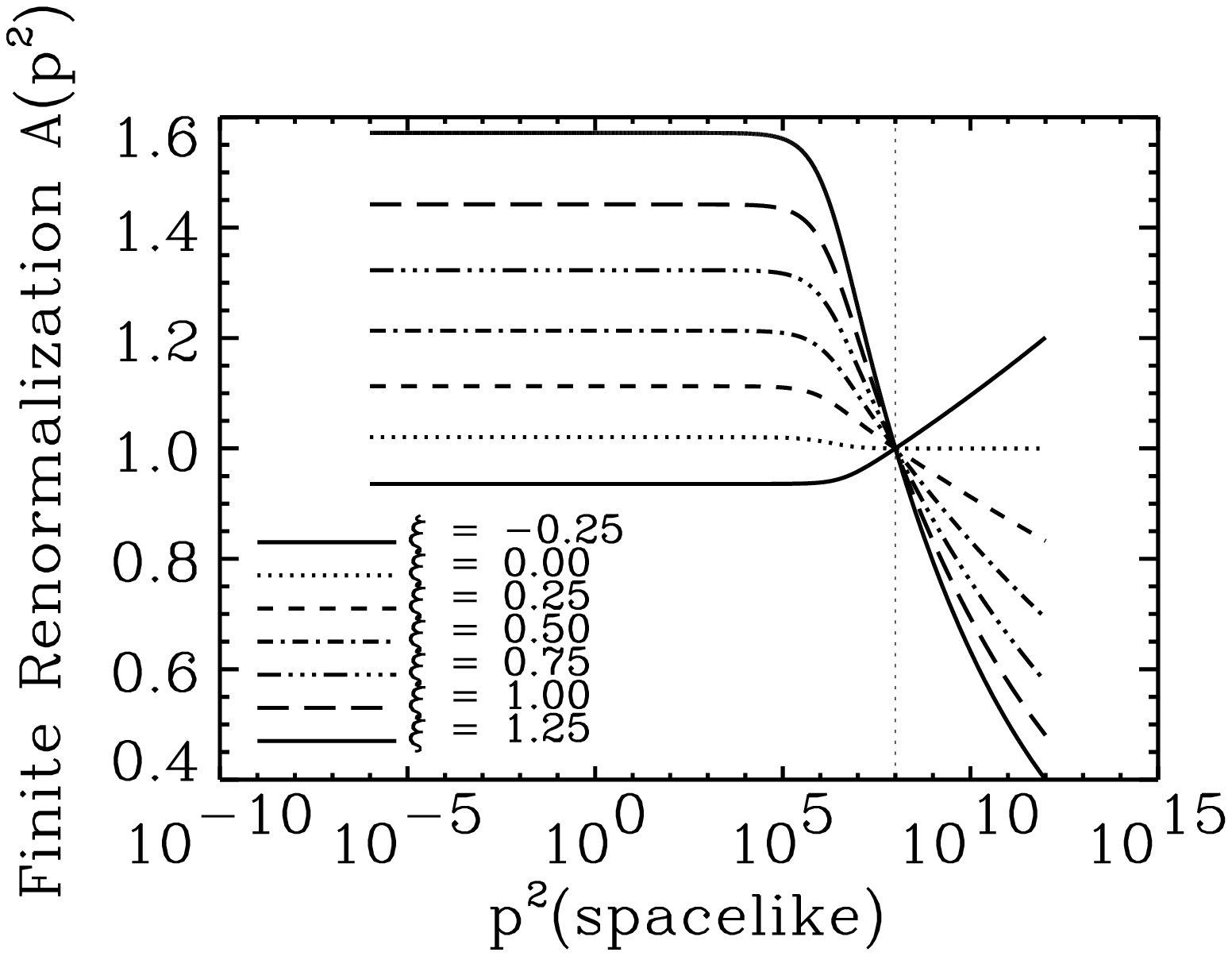}
    \vspace{0.5cm}
  \setlength{\epsfxsize}{12.0cm}
  \centering
      \epsffile{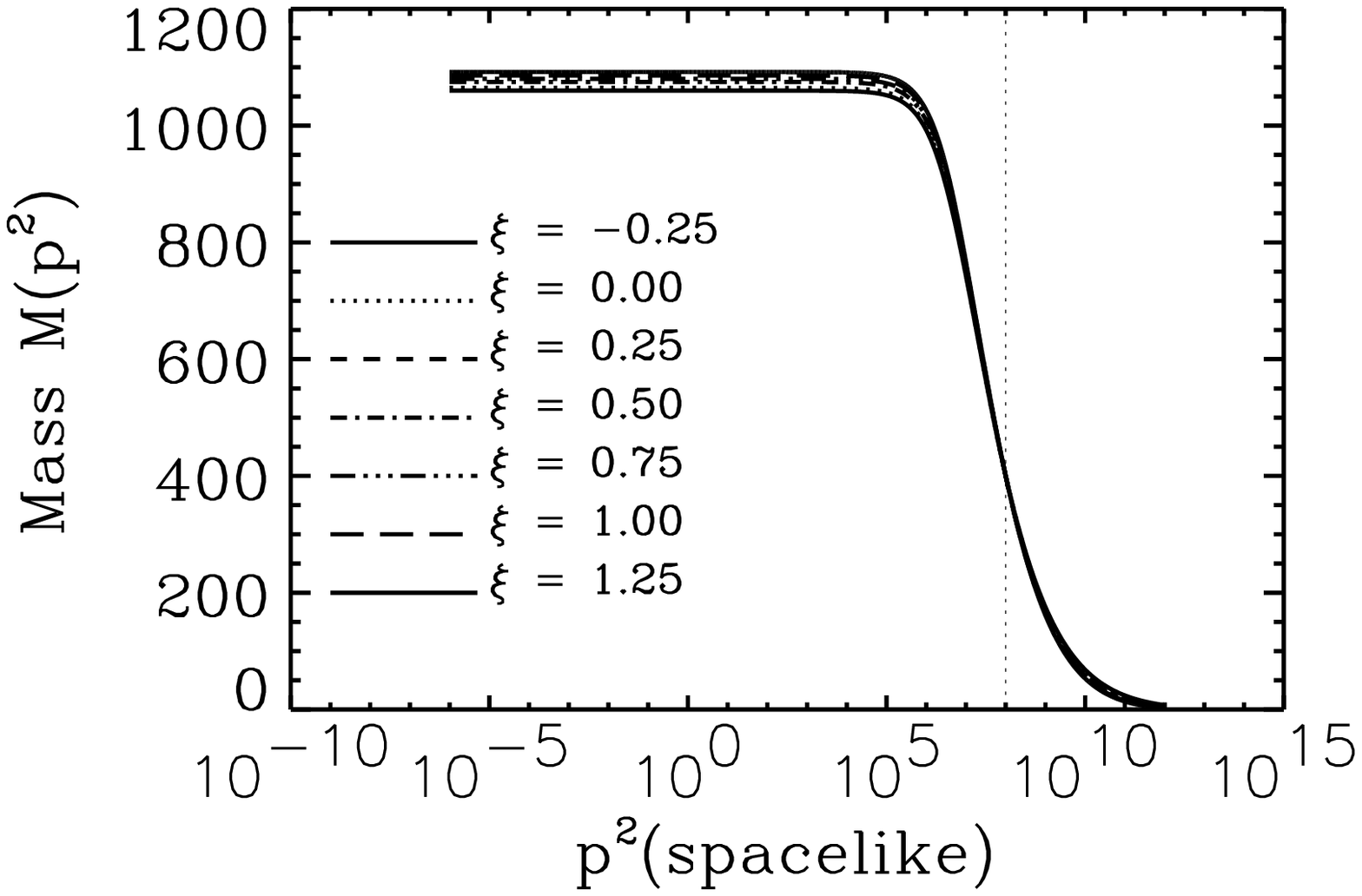}
  \parbox{130mm}{\caption{
      The finite renormalization $A(p^2)$ and the mass function $M(p^2)$
      are shown for various gauge parameters $\xi$.  These results have
      coupling $\alpha=1.00$, renormalization point $\mu^2=10^8$,
      and renormalized mass $m(\mu)=400$.  In the low $p^2$ region
      the larger gauge parameter has the larger value of $M(p^2)$.
  \label{diff_gauges}}}
  \vspace{0.5cm}
\end{figure}

Fig.~\ref{diff_gauges} shows a family of solutions
characterized by
  $\alpha=1.00$, $\mu^2 = 1 \times 10^8$, $m(\mu) = 400$,
and gauge parameters from -0.25 to 1.25.
We see that while $A$ and $B$ are strongly gauge dependent,
the mass function $M(p^2)\equiv B(p^2)/A(p^2)$
is relatively insensitive to $\xi$.
The location of the mass pole of the
physical electron must of course be independent of gauge,
and this gauge independence has been demonstrated explicitly
using the WTI for example by Atkinson and Fry \cite{Atk+Fry}.
Their proof assumes that the bare mass $m_0(\Lambda)$
is itself independent of gauge.
Hence, in a fully gauge covariant treatment the mass function is independent
of gauge at two scales (i.e., at the mass pole and at the UV regularization
scale $\Lambda$).  In our study we find that the 
mass function is relatively insensitive to the choice of gauge for all
$p^2$.  The nature of the Landau-Khalatnikov transformations
\cite{LKTF} makes the possibility of a gauge independent $M(p^2)$
seem rather unlikely.

\begin{figure}[htb]
  \setlength{\epsfxsize}{12.0cm}
  \centering
      \epsffile{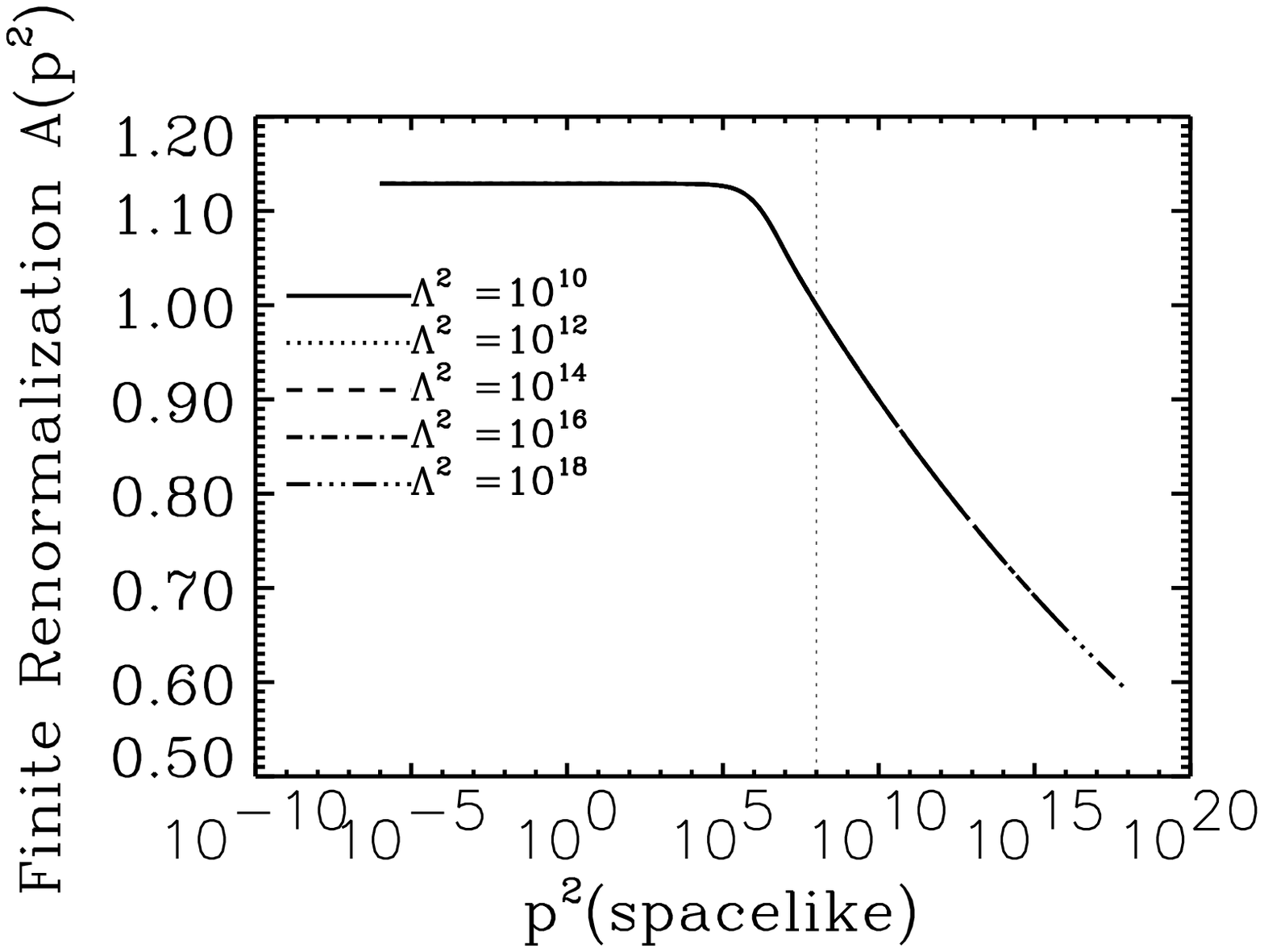}
    \vspace{0.5cm}
  \setlength{\epsfxsize}{12.0cm}
  \centering
      \epsffile{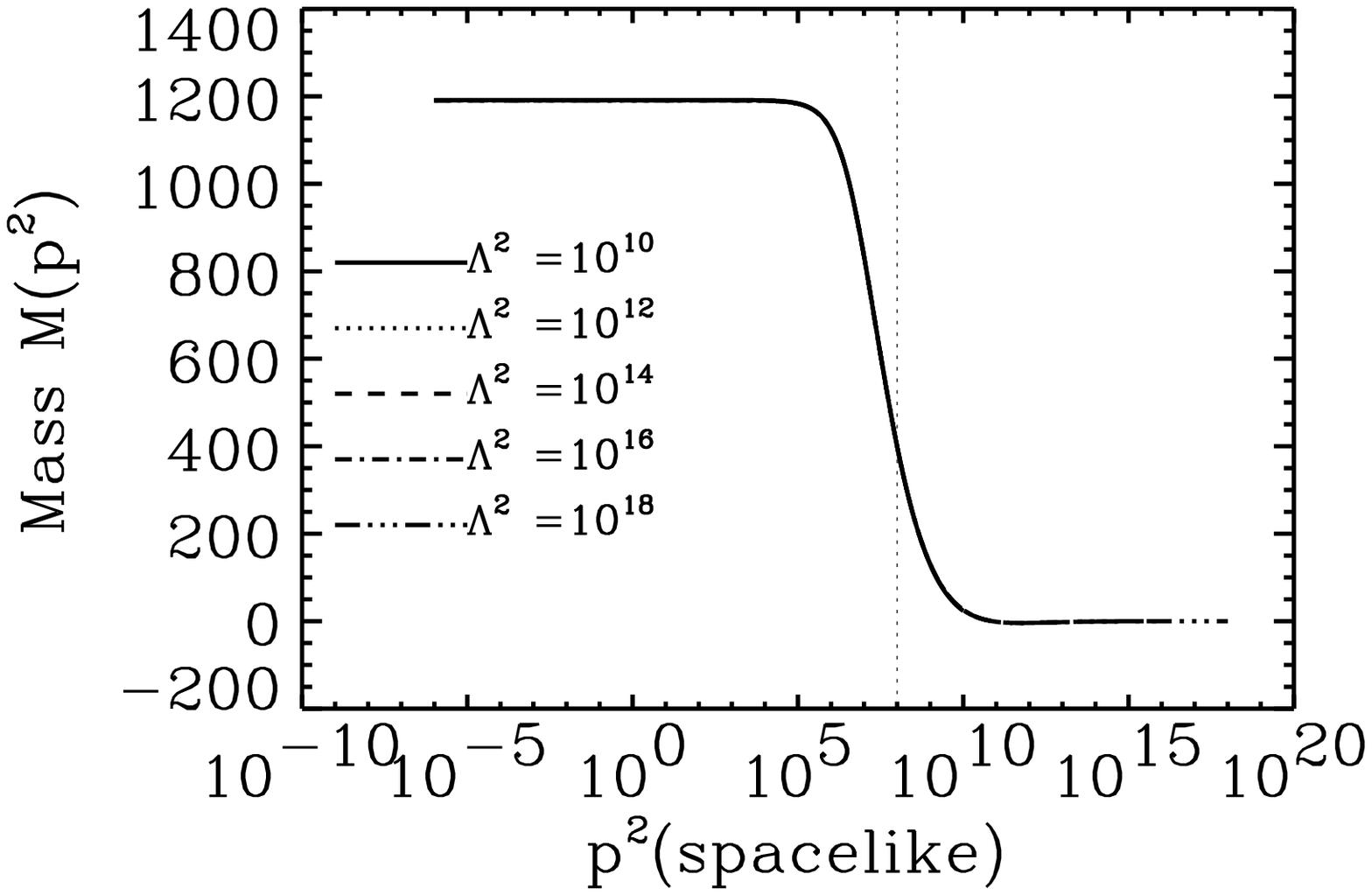}
  \parbox{130mm}{\caption{
      The finite renormalization $A(p^2)$ and the mass function $M(p^2)$
      are shown for various choices of the regularization parameter
      (i.e., ultraviolet cut-off) $\Lambda$.  These results have
      coupling $\alpha=1.15$, renormalization point $\mu^2=10^8$,
      renormalized mass $m(\mu)=400$, and gauge parameter $\xi=0.25$.
      The stability of the
      subtractive renormalization procedure is apparent.
  \label{diff_UV}}}
  \vspace{0.5cm}
\end{figure}

The stability of the renormalized DSE solutions
with respect to variations in the ultraviolet cutoff
is evident in Fig.~\ref{diff_UV}.
This graph shows solutions with $\alpha = 1.15$, $\mu^2 = 10^8$,
$m(\mu) = 400$, and gauge $\xi=0.25$.  The cutoff $\Lambda^2$ was
varied over several orders of magnitude with no apparent change in
the solutions over the common range of momenta.
This numerical stability was shown in other tests as well.
For instance, we extracted the mass $M(p^2 = 0)$ for solutions with
$\alpha=1.00$, $\mu^2 = 10^4$, $m(\mu) = 0$, and observed variations
of less than one part in $10^4$ as the UV cutoff was varied over six
orders of magnitude.

Tables \ref{Landau_tbl}, \ref{gp025_tbl}, and \ref{gp050_tbl}
show the evolution of the renormalization constants
$Z_2(\mu,\Lambda)$, $Z_m(\mu,\Lambda)$, and the cutoff-dependent
bare mass $m_0(\Lambda)$ as a function of the UV regulator $\Lambda$.
We see that as we move further from Landau gauge $Z_2(\mu,\Lambda)$
decreases more rapidly with increasing $\Lambda$.  In addition,
we observe that the bare mass exhibits decaying oscillations with increasing
$\Lambda$, which is directly related to the oscillations characteristic
of the supercritical coupling and subtractive renormalization at large
$p^2$ (see additional discussion later). 

\begin{figure}[htb]
  \setlength{\epsfxsize}{12.0cm}
  \centering
      \epsffile{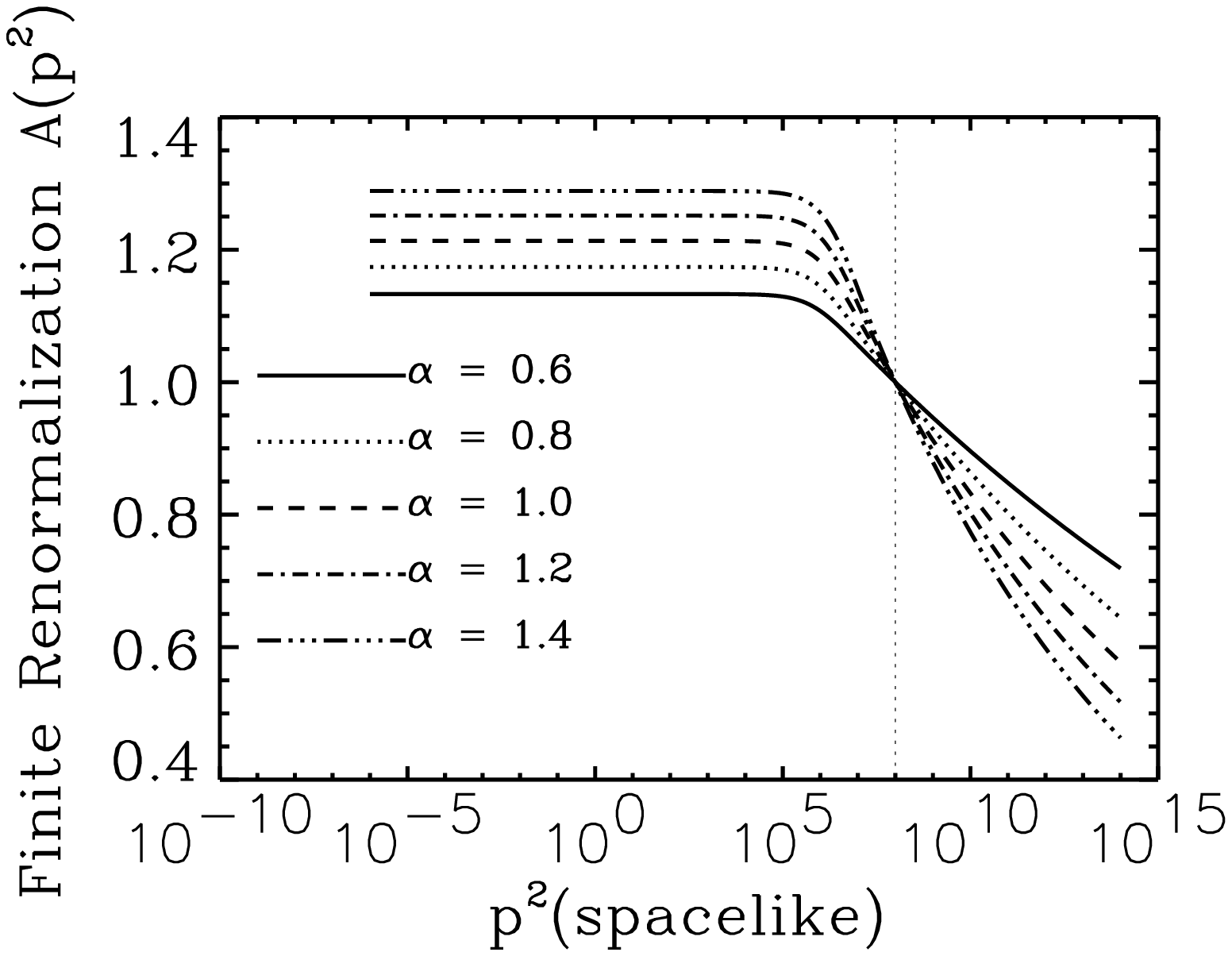}
    \vspace{0.5cm}
  \setlength{\epsfxsize}{12.0cm}
  \centering
      \epsffile{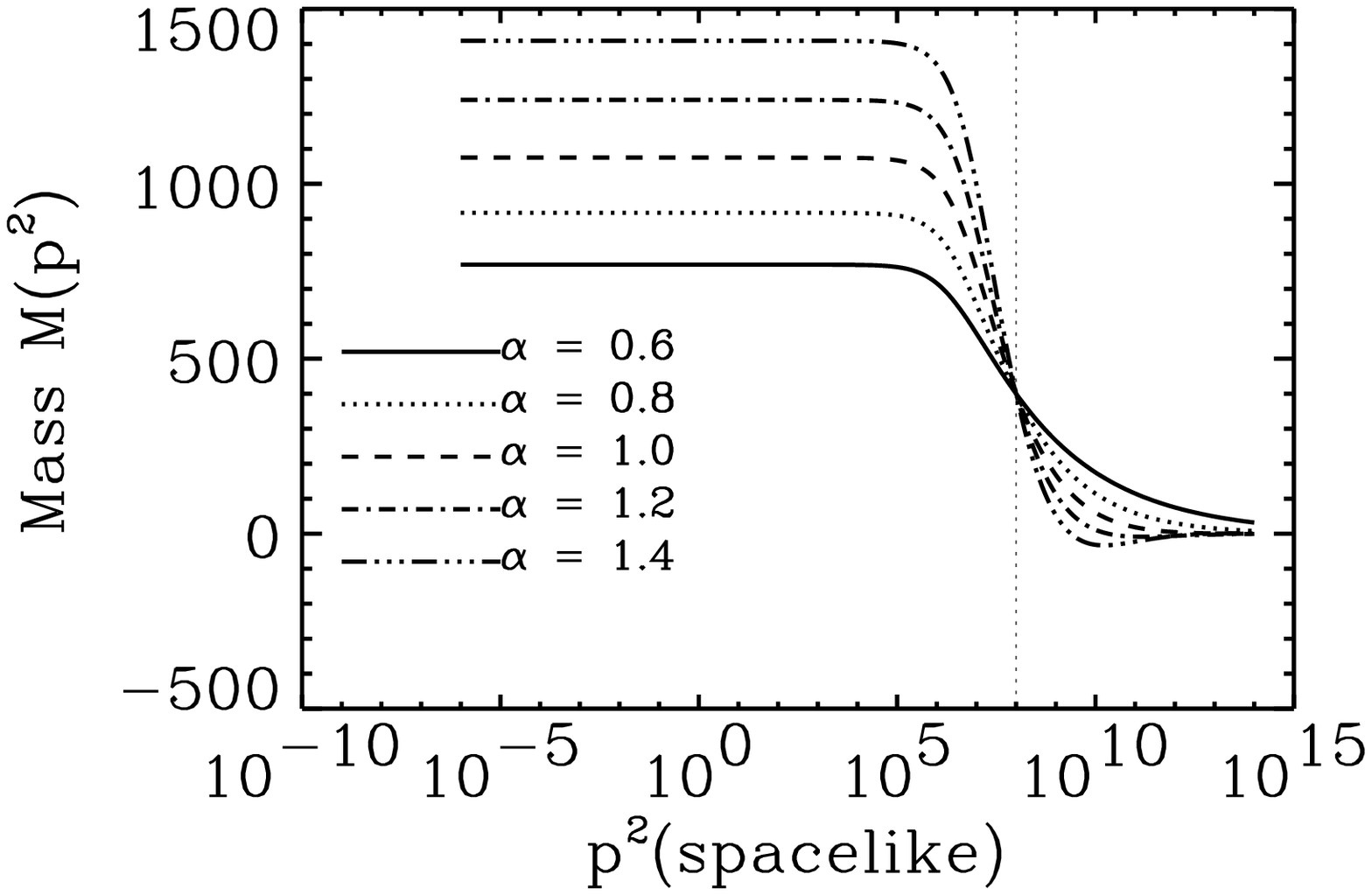}
  \parbox{130mm}{\caption{
      The finite renormalization $A(p^2)$ and the mass function $M(p^2)$
      are shown for various choices of the coupling strength $\alpha$.
      These results have renormalization point $\mu^2=10^8$,
      renormalized mass $m(\mu)=400$, and gauge parameter $\xi=0.50$.
  \label{diff_alpha}}}
  \vspace{0.5cm}
\end{figure}

Fig.~\ref{diff_alpha} shows solutions with the coupling varying from
subcritical ($\alpha = 0.6$) to supercritical ($\alpha = 1.4$) values,
with identical renormalization point, renormalized mass and gauge.
Here we see that the the nodes in the mass function
$M(p^2)$ move to lower momenta and the oscillations become more pronounced
as the coupling is increased further and further above critical coupling.

In order to test the gauge-invariance of the chiral critical point,
we extracted the critical coupling from solutions in the Landau gauge
and in two other covariant gauges, with $\xi = 0.25$ and 0.5.
Miransky {\it et al.}\/, \cite{FGMS,Miransk1}
working in the quenched ladder approximation in Landau gauge,
found that the infrared limit of the dynamical mass
$M(0)$ has an infinite-order phase transition,
\begin{equation}
  M(0) \simeq 4 \Lambda \exp\left[
        - \frac{\pi}{ \sqrt{ (\alpha/\alpha_c) - 1 } }
			   \right]\:.
\end{equation}
Following their approach, we assume a similar form
for the dynamical mass near the critical coupling,
\begin{equation}  \label{crit_functionalform}
  M(0) = M \exp \left[
	   - \frac{c}{\left( (\alpha/\alpha_c) - 1 \right)^\beta}
	        \right] \:,
\end{equation}
and construct an order parameter which is
expected to have a second-order phase transition,
$-1/\log[ M(0) / M']$.
Since the inherent mass scale $M$ is not known {\it a priori\/},
it is necessary to choose a reasonable scale $M'$,
and then calculate a corrected fit,
which also yields the actual value of $M$.

\begin{figure}[htb]
  \setlength{\epsfxsize}{12.0cm}
  \centering
      \epsffile{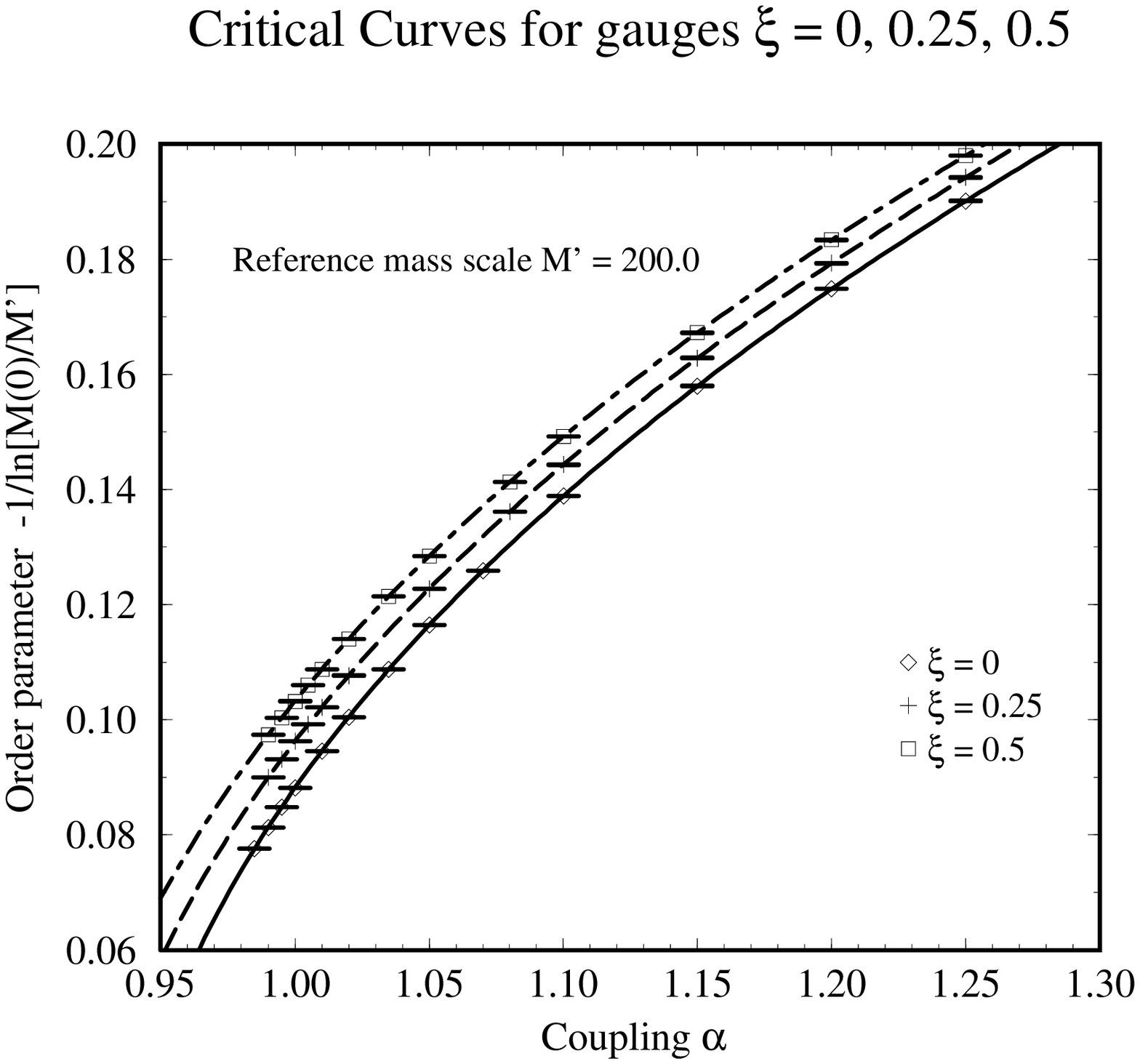}
  \parbox{130mm}{\caption{
       The critical curves for three choices of gauge parameter
       showing the existence of residual gauge-dependence in
       the Curtis-Pennington vertex.  For the purposes of studying the
       chiral phase transition all solutions were
       renormalized with the choice, $m(\mu) = 0$\/,
       with the renormalization point $\mu^2 = 10^4$.
       The order parameter is evaluated using an arbitrary reference
       mass scale choice of $M' = 200.0$.
       Diamonds ($\Diamond$) connected by the solid smooth curve,
       are order parameter values for the Landau gauge;
       pluses ($+$) connected by the dashed smooth curve,
       are values for $\xi=0.25$;
       and boxes ($\Box$) connected by the dot-dashed smooth curve,
       are values for $\xi=0.5$\/.
  \label{crit_curves}}}
  \vspace{0.5cm}
\end{figure}

DSE solutions were obtained for several values of the coupling
in each gauge, with the renormalization $\mu^2=10^4$.  We set
$m(\mu)=0$ for the purpose of studying the transition.
The IR mass limit was extrapolated for each solution
and then the order parameters were calculated
using an assumed mass scale $M' = 200$.
The resulting critical curves are shown
along with the $M(0)$ values from the solutions,
in Fig.~\ref{crit_curves}.
The parameters from the nonlinear fits are given
in Table \ref{crit_fits_tbl}.
The critical exponents $\beta$ are the same to within
their numerical tolerance, and suggest that $\beta$
may be independent of gauge.
Although the values of $\alpha_c$ are close in value,
there is clear evidence of residual gauge-dependence.

Our value for $\alpha_c$ in the Landau gauge is very close
to the value of 0.933667 found by
Atkinson {\it et al}\/ \cite{ABGPR} in their
bifurcation analysis of the solutions of the fermion DSE
with the Curtis-Pennington vertex.  In addition we find that the
critical coupling varies with $\xi$ in the same direction.
Recall that these authors used the unrenormalized equations and
relaxed the UV momentum cutoff to infinity
in order to remove cutoff artifacts, whereas we have used
subtractive renormalization and our gauge covariance correction.
Hence we can anticipate a small difference between the critical
couplings in our approaches.

\begin{figure}[t]
  \setlength{\epsfxsize}{12.0cm}
  \centering
      \epsffile{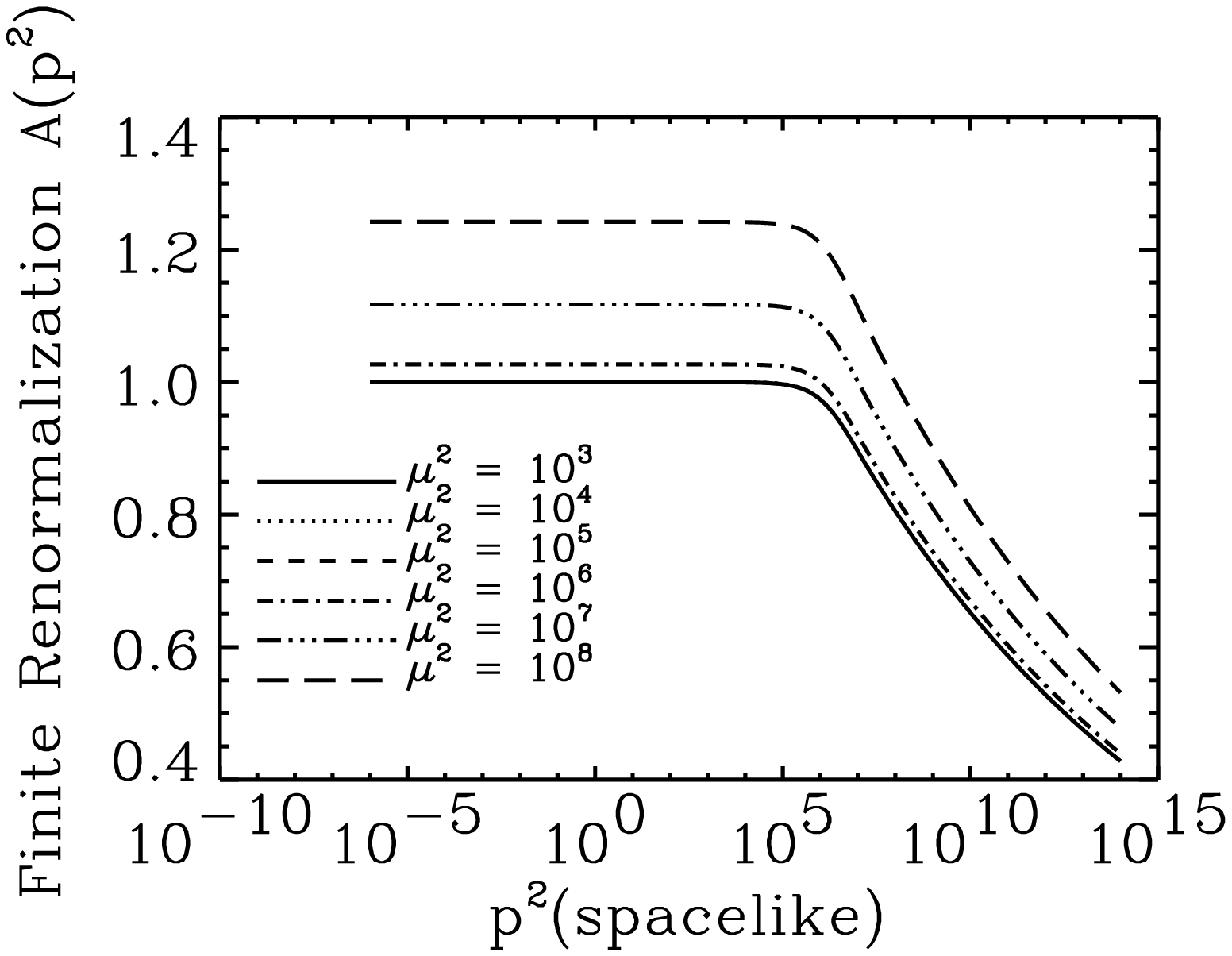}
    \vspace{0.5cm}
  \setlength{\epsfxsize}{12.0cm}
  \centering
      \epsffile{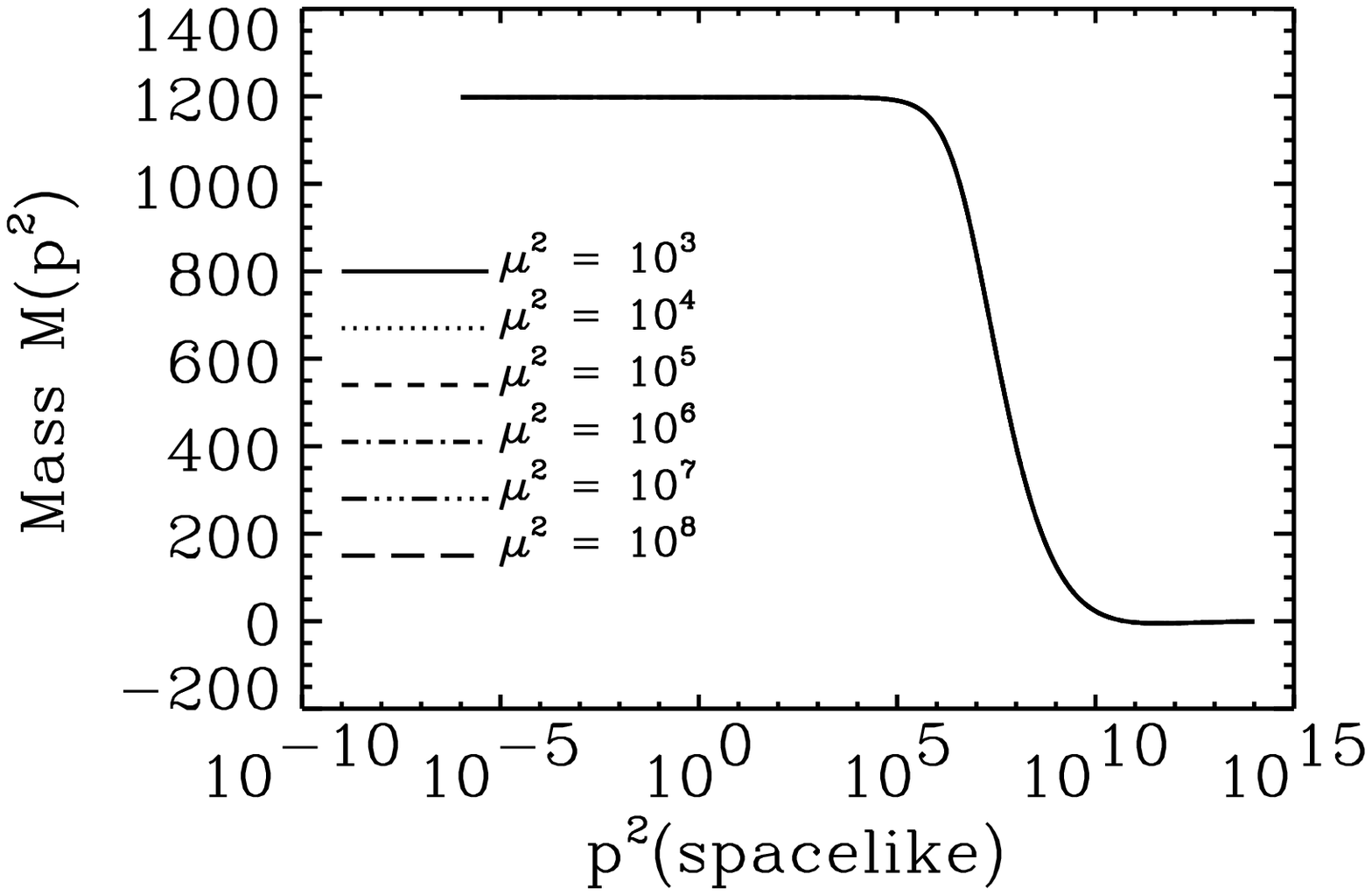}
  \parbox{130mm}{\caption{
      The finite renormalization $A(p^2)$ and the mass function $M(p^2)$
      are shown for various choices of renormalization point.
      These results have coupling strength $\alpha=1.15$ and
      gauge parameter $\xi=0.50$.  Each of these results
      corresponds to $M(p^2)=400$ at $p^2=10^8$.  Hence, $M(p^2)$ is 
      renormalization point independent and $A(p^2)$ varies as described
      in Eq.\ (\protect\ref{ren_pt_transf}).
  \label{diff_mu}}}
  \vspace{0.5cm}
\end{figure}

Fig.~\ref{diff_mu} shows a family of equivalent solutions
renormalized at different momentum scales $\mu$ and these provide a direct
check on the behavior predicted in Eq.~(\ref{ren_pt_transf}).
All have coupling $\alpha=1.15$ and $\xi=0.5$, and the renormalization
scale $\mu^2$ is stepped by powers of 10, with renormalized masses
$m(\mu)$ chosen such that $m(\mu)=M(\mu^2)$ for each $\mu$.
It is clear that the resulting mass curves are identical at all $p^2$
as expected, and in each case $A(\mu;p^2)$ scales as predicted in
Eq.~(\ref{ren_pt_transf}) showing that the renormalized DSE does transform
correctly.

\begin{figure}[tb]
  \setlength{\epsfxsize}{12.0cm}
  \centering
      \epsffile{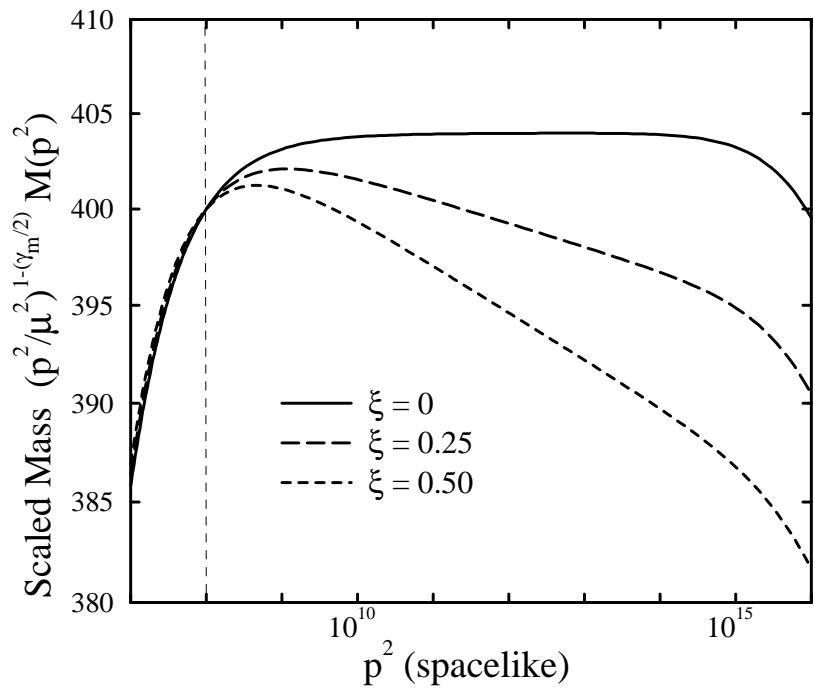}
  \parbox{130mm}{\caption{
     Asymptotic mass scaling below critical coupling
      as a function of gauge,
      with $\alpha=0.5$, $\mu^2 = 10^8$, $m(\mu) = 400$.
      The scaling applied to the masses uses the anomalous
      dimension found for the Landau guage,
        $\gamma_m(\xi=0) = 1.716638$.
      the extracted anomalous dimensions for the other
      two curves are
        $\gamma_m(0.25)  = 1.713498$, and
        $\gamma_m(0.5)   = 1.711274$.
  \label{scaled_mass}}}
  \vspace{0.5cm}
\end{figure}

The anomalous dimension of the mass, $\gamma_m$, is defined by
the asymptotic scaling of the dynamical mass with $p^2$,
\begin{equation}
  M(p^2) \sim \left( \frac{p^2}{\mu^2} \right)^{(\gamma_m/2) - 1}\:.
\end{equation}
As well as depending on the coupling, it shows a slight dependence
on the gauge, as shown in Fig.~\ref{scaled_mass}.
The dynamical masses shown are from DSE solutions with
$\alpha=0.5$, $\mu^2 = 10^8$, and $m(\mu) = 400$,
in Landau gauge and in gauges $\xi=0.25$ and 0.5.
They are scaled by multiplication
with $(p^2/\mu^2)^{1-(\gamma_m/2)}$,
where the value of $\gamma_m$ used was 
that extracted from the Landau gauge solution,
$\gamma_m = 1.716638$\/.
The gauge-dependence of $\gamma_m$ shows up as the slight difference
in slopes on the log-log plot.
(The dips apparent at the end of the curves are due to
having a hard momentum cut-off $\Lambda^2=10^{16}$.
As $\Lambda$ is increased these move to higher momenta also.)
In Landau gauge, the actual power of $1/p^2$ with which $M(p^2)$
falls asymptotically is $s = 1 - (\gamma_m/2) = 0.141681$\/.
For the gauges $\xi=0.25$ and 0.5, the anomalous dimensions
are 1.713948 and 1.711274,
giving powers of $1/p^2$ equal to 0.143026 and 0.144363 respectively.

Miransky has studied the form of the mass renormalization $Z_m$
in the bare vertex approximation in Landau gauge, and without subtractive
renormalization \cite{Miransk2,Miransk3}.
In this treatment he finds
  $Z_m(\mu,\Lambda) = (\mu^2/\Lambda^2)^{\frac{1}{2} - \gamma'}$\/,
with the exponent
\begin{equation}  \label{gamma_prime}
  \gamma'(\alpha) = \frac{1}{2} \sqrt{1 - \alpha/\alpha_c}\:,
\end{equation}
where the critical coupling for DCSB in that approximation
is $\alpha_c = \pi/3$.
This would imply an asymptotic scaling for the dynamical mass
that goes like $M(p^2) \sim (p^2)^{\gamma' - 1/2}$,
so that the anomalous mass dimension would be related to $\gamma'$
by $\gamma' = (\gamma_m - 1)/2$\/.
Recent articles by Holdom \cite{Holdom}
and Mahanta \cite{Mahanta1,Mahanta2}
claim that for quenched theories {\em at criticality}\/
the mass anomalous dimension $\gamma_m$ should be exactly 1,
giving $M(p^2) \sim 1/p$ as in the bare approximation,
and that in particular
this result should be independent of the gauge \cite{Mahanta2}.
In Miransky's treatment this corresponds to the vanishing of $\gamma'$ at
the critical coupling.  We find for subcritical couplings
that $\gamma_m\simeq 1.71$ and that it is
gauge insensitive.  Hence, as for the pure CP-vertex case,
our gauge corrected CP vertex does not lead to $\gamma_m=1$.

\begin{figure}[tb]
  \setlength{\epsfxsize}{12.0cm}
  \centering
      \epsffile{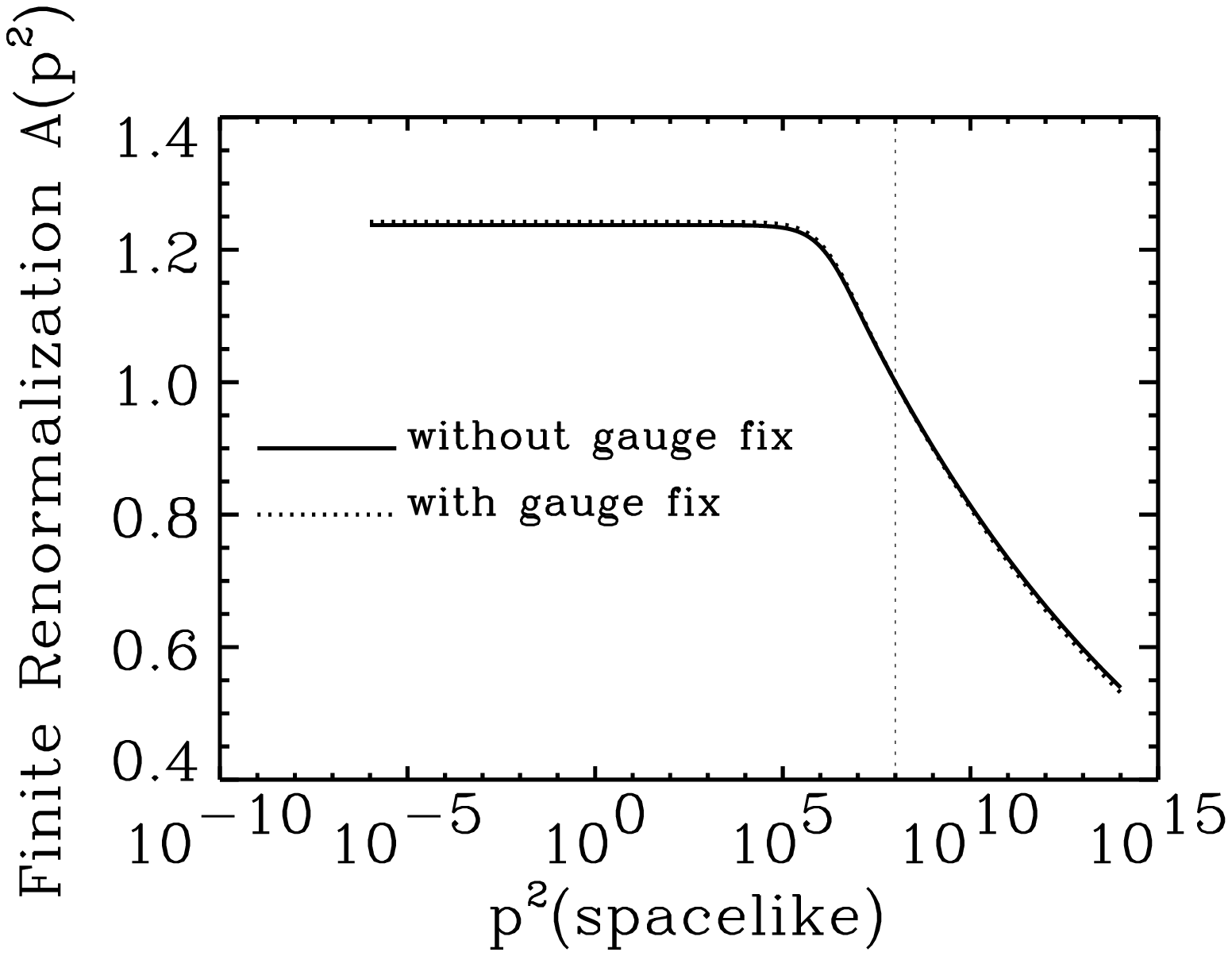}
    \vspace{0.5cm}
  \setlength{\epsfxsize}{12.0cm}
  \centering
      \epsffile{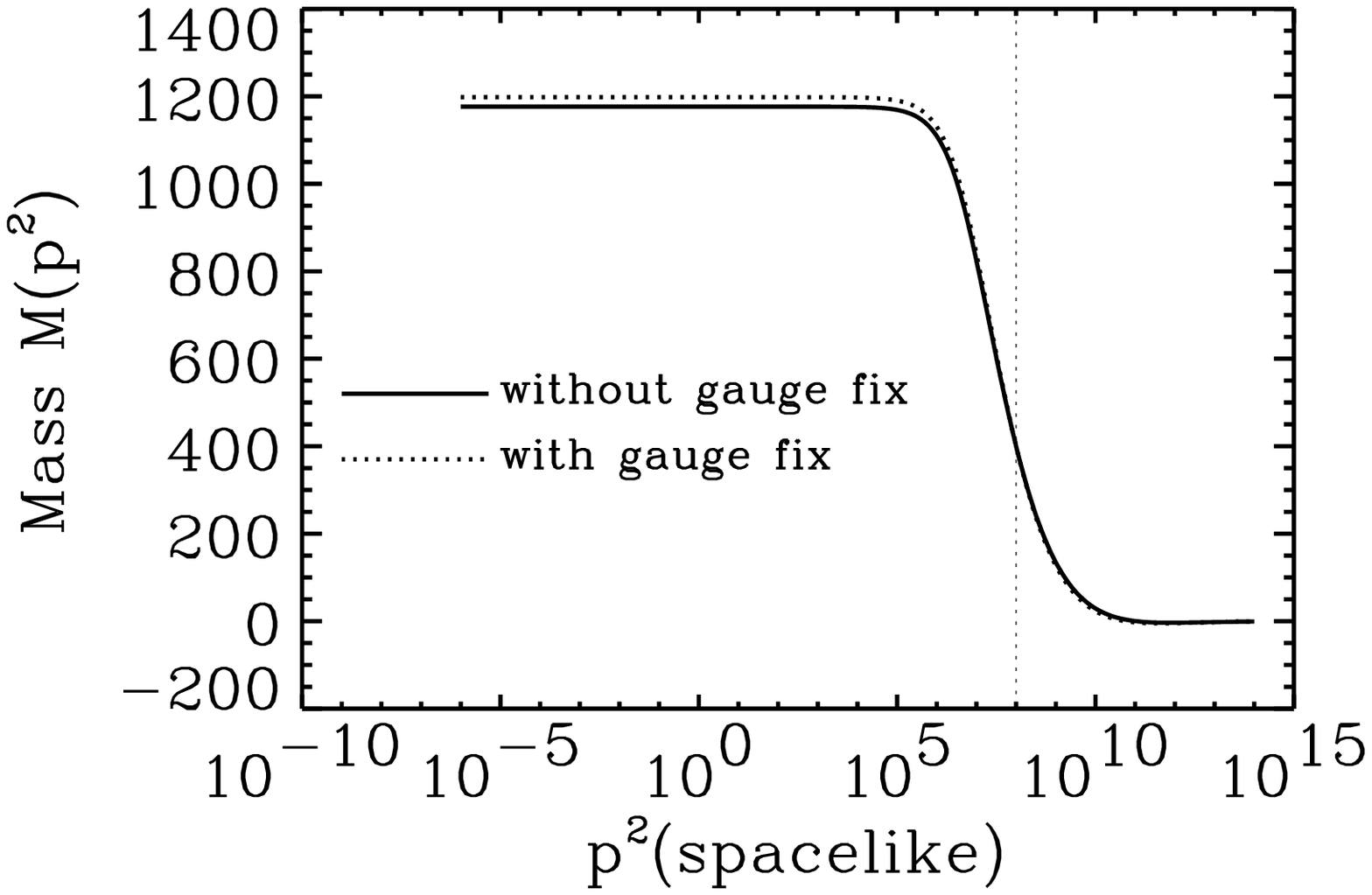}
  \parbox{130mm}{\caption{
      The finite renormalization $A(p^2)$ and the mass function $M(p^2)$
      are shown with and without the gauge covariance correction.
      We see that the correction is a relatively small effect.
      These results have coupling strength $\alpha=1.15$,
      renormalization point $\mu^2=10^8$,
      renormalized mass $m(\mu)=400$, and gauge parameter $\xi=0.50$.
  \label{gauge_cov_fix}}}
  \vspace{0.5cm}
\end{figure}

The gauge covariance correction described in the Appendix leads to an exact
restoration of gauge covariance in the subcritical case with no explicit
chiral symmetry breaking, i.e., no bare mass
(see, e.g., Ref.~\cite{dongroberts}).
The difference between a DSE solution with naive cutoff
regularization and those with the gauge covariance correction
is shown in Fig.~\ref{gauge_cov_fix}.
Both solutions have $\alpha=1.15$\/, $\xi=0.5$\/, and are
renormalized at  $\mu^2 = 10^8$ and $m(\mu) = 400$\/.
The quantitative change induced by our gauge covariance
correction was found to be relatively small in the presence of a
substantial mass function $M(p^2)$.

\begin{figure}[htb]
  \setlength{\epsfxsize}{12.0cm}
  \centering
      \epsffile{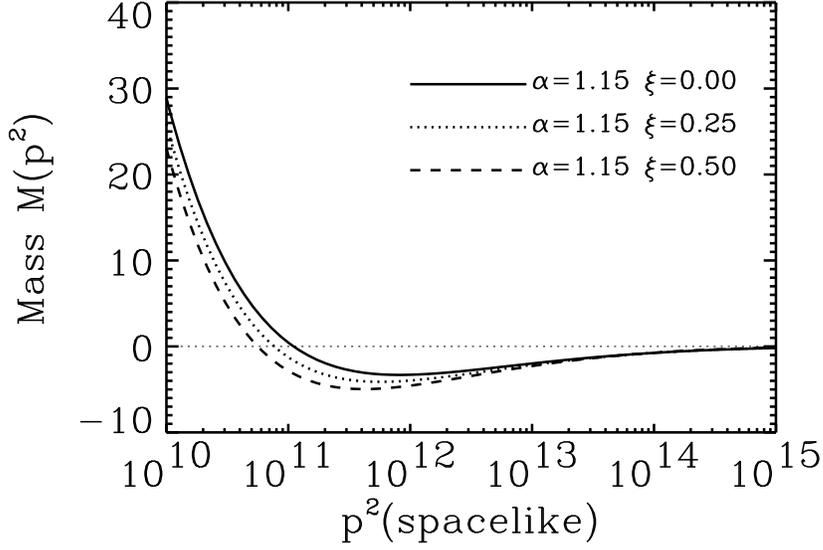}
  \parbox{130mm}{\caption{
      Detail of the node in the mass function $M(p^2)$
      for various gauge choices.
      These results have coupling strength $\alpha=1.15$,
      renormalization point $\mu^2=10^8$,
      and renormalized mass $m(\mu)=400$.
  \label{mass_nodes}}}
  \vspace{0.5cm}
\end{figure}

\begin{figure}[htb]
  \setlength{\epsfxsize}{10.0cm}
  \centering
      \epsffile{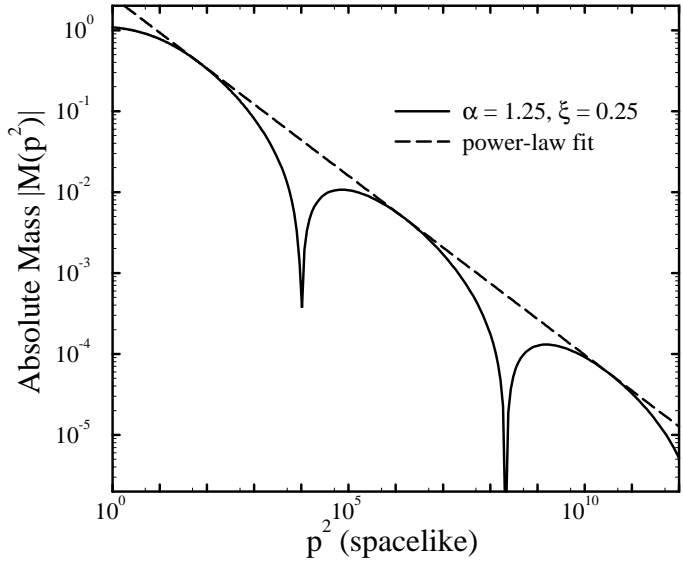}
  \parbox{130mm}{\caption{
      Absolute value of the dynamical mass,
      showing damped oscillations periodic in $\ln(p^2)$.
      The solution shown has $\alpha= 1.25$, $\xi= 0.25$,
      and is renormalized with $\mu^2=10^4$, $m(\mu) = 0$.
      The power-law fit, which runs tangent to the dynamical
      mass curve, is
      $C (p^2/\mu^2)^{(\gamma_m/2) - 1}$\/,
      with $C = 4.394 \times 10^{-2}$, $\gamma_m = 1.115$\/.
  \label{mass_damped_osc}}}
  \vspace{0.5cm}
\end{figure}

We find, as in our previous study in Landau gauge \cite{qed4_hw},
that for supercritical couplings the dynamic mass 
crosses zero; for solutions away from the chiral
limit, the position of the first node depends on the gauge,
as shown in Fig.~\ref{mass_nodes}.
In fact as the cutoff is increased $M(p^2)$ shows damped oscillations
periodic in $\log{p^2}$\/, as shown in Fig.~\ref{mass_damped_osc}.
This has been discussed by several authors
\cite{Miransk2,Mahanta2,pirho_fr};
in particular it is shown
using some simplifying approximations, in \cite{pirho_fr}
that in Landau gauge, the DSE reduces to a differential equation
for $M(p)$ which has the solution
\begin{equation}   \label{mass_osc_form}
  M(p) = \kappa \left( \frac{p}{\mu} \right)^{-1}
	 \cos \left(
	   \frac{1}{2} \ln(p^2/\mu^2) \sqrt{\alpha/\alpha_c - 1} + \phi
	     \right)\:,
\end{equation}
with $\kappa \cos \phi = m(\mu)$\/.
However, the approximations used in deriving this result
are not applicable outside Landau gauge and even in Landau gauge
lead to differences from the present treatment.
We find that the functional form is substantially correct, but that
both the mass dimension and the period of oscillations depend on
the coupling and the choice of gauge.
In fact, for the case shown in Fig.\ \ref{mass_damped_osc},
the mass dimension that fits $|M(p^2)|$ is $\gamma_m = 1.115$\/.
The dependence of the period on $\alpha$ and $\alpha_c$
is also not as simple as that in Eq. (\ref{mass_osc_form}).

\section{Summary and Conclusions}
\label{sec_conclusions}

We have extended our previous work on the numerical renormalization
of the DSE \cite{qed4_hw} to arbitrary covariant gauges.
The procedure is straightforward to implement and extremely stable.
It becomes numerically more challenging
for covariant gauges far removed from Landau gauge and for large
couplings ($\alpha\gg 1$). The importance of the approach is that it
removes the issue of cut-off dependence and allows solutions to be obtained
for any choice of renormalization point.
We have described the procedure for performing renormalization
group transformations between solutions
with different renormalization points.  We saw that a knowledge
of the solution at one renormalization point automatically provides
the solutions at all renormalization points.

This then allows also comparisons with results from lattice studies of QED,
which should prove useful in providing further guidance in the choice
of reasonable {\it Ans\"atze} for the vertex and photon propagator. 
Without renormalization only the unrenormalized, regulated quantities
would be be obtained and any such comparisons would be meaningless.
In addition,
in order to study the nonperturbative behaviour
of renormalization constants such as
$Z_1(\mu,\Lambda)$, $Z_2(\mu,\Lambda)$, and $Z_m(\mu,\Lambda)$
they must be numerically extracted and so a method
such at that described here would be essential.

The context of this study has been quenched four-dimensional QED
with a modified Curtis-Pennington vertex,
since that vertex {\it Ansatz}\/ has the desirable properties
of making the solutions approximately gauge-invariant
and also
multiplicatively renormalizable up to next-to-leading log order.
The technique described can be generalized to
apply elsewhere (e.g., QCD), whenever numerical
renormalization is required.

The solutions are stable
and the renormalized quantities become independent of regularization
as the regularization is removed, which is as expected.
For example, the mass function $M(p^2)$ and the
momentum-dependent renormalization $A(p^2)\equiv 1/Z(p^2)$ are unchanged 
to within the numerical accuracy of the computation as the
integration cutoff is increased by many orders of magnitude.
The mass renormalization constant $Z_m(\mu,\Lambda)$
converges to zero with increasing $\Lambda$
because the mass function $M(p^2)$ falls to zero
sufficiently rapidly at large $p^2$.
The absence of divergences of $Z_1(\mu,\Lambda)=Z_2(\mu,\Lambda)$,
$m_0(\Lambda)$, and $Z_m(\mu,\Lambda)$ in the limit $\Lambda\to\infty$
is a purely nonperturbative result
and is in sharp contrast to the
perturbative case where these constants diverge at all orders.

In order to study the critical point and exponents
for the transition to DCSB,
one sets the renormalized mass $m(\mu)$ to 0 and varies the coupling.
For subcritical couplings $m_0(\Lambda)$ remains zero, and the
dynamical mass $M(p^2)$ is identically zero, while for supercritical
couplings $M(p^2)\neq 0$ [and $m_0(\Lambda)\neq 0$ for finite $\Lambda$].
We have extracted the critical coupling for DCSB in Landau gauge ($\xi=0$),
and in gauges with $\xi=0.25$ and 0.5.  Our Landau gauge result is
very close to the value found in \cite{ABGPR}; the values in the
other gauges show a small residual gauge-dependence.

For subcritical couplings,
we find that the mass renormalization
$Z_m(\mu,\Lambda)$ scales approximately as 
$Z_m(\mu,\Lambda) \propto
    (\mu^2/\Lambda^2)^{1 - \gamma_{m}(\alpha,\xi)/2}$,
where e.g. $\gamma_{m}(0.5,0)= 1.716638$. 
Above the critical coupling, the mass function shows damped
oscillations around zero, periodic in $\ln(p^2)$.
We have extracted mass anomalous dimensions $\gamma_m$
for some subcritical and supercritical couplings, and find
them all greater than 1.

We have shown that our modified Curtis-Pennington vertex, while
removing the violation of gauge covariance in the massless
subcritical case (i.e., when there is no explicit or dynamical chiral
symmetry breaking), has not been sufficient to remove the small
residual violation of gauge covariance in the general case.
Hence it is important to attempt to extend
this work to include other regularization
schemes (e.g., dimensional regularization) and to
vertices of the Bashir-Pennington type \cite{BP1,BP2}.

\begin{acknowledgements}
We thank Michael Neuling for assistance with some of the numerical
calculations.
This work was partially supported by the Australian Research Council,
by the U.S. Department of Energy
through Contract No.\ DE-FG05-86ER40273,
and by the Florida State University Supercomputer Computations Research
Institute which is partially funded by the Department of Energy
through Contract No.\ DE-FC05-85ER250000.
The work of CDR was supported by the US Department of Energy,
Nuclear Physics Division, under contract number W-31-109-ENG-38.              
This research was also partly supported by grants of supercomputer time
from the U.S. National Energy Research Supercomputer Center
and the Australian National University Supercomputer Facility.
AGW thanks the (Department of Energy) Institute for Nuclear Theory at the
University of Washington for its hospitality and partial support during
the completion of this work.
\end{acknowledgements}

\begin{appendix}

\section{UV Regulator and Gauge Covariance}
\label{appdx_a}

Regulators applied to divergent integrals in field theory
always destroy some continuous symmetry, and in particular
the use of a momentum cutoff destroys gauge covariance.
This appendix describes a modification of the
self-energy integrals in the regularized DSE,
which will at least partially restore this symmetry.

The basis of this change in the regularization scheme is that
when the self-energy $\Sigma$, given in Eq. (\ref{reg_Sigma}),
is evaluated under cutoff regularization, it contains a term
related to the vertex WTI, which should vanish
but which integrates to give a nonzero contribution
because the cutoff regularization scheme
is not translationally invariant \cite{dongroberts,BP1}.
We therefore evaluate this term separately
and subtract it from the self-energy.
It turns out to affect only the value of the
renormalization $Z_2(\mu,\Lambda)$.

We write the photon propagator, in Minkowski momentum space and in an
arbitrary covariant gauge, as
\begin{eqnarray}
  D_{\mu \nu}(q) & = & \left( -g_{\mu\nu} + \frac{q_\mu q_\nu}{q^2} \right)
	                 \frac{1}{1+\Pi(q^2)} \frac{1}{q^2} 
		       - \xi \frac{q_\mu q_\nu}{q^2} \frac{1}{q^2}
  \\
  & \equiv & -{\cal T}_{\mu\nu}(q) \frac{1}{1+\Pi(q^2)} \frac{1}{q^2} 
	     - {\cal L}_{\mu \nu}(q) \xi \frac{1}{q^2}
  \\
  & \equiv & D^{\rm T}_{\mu \nu}(q) + D^{\rm L}_{\mu \nu}(q)\:.
\end{eqnarray}
where ${\cal T}_{\mu \nu}(q)$ and ${\cal L}_{\mu \nu}(q)$ are the transverse
and longitudinal projectors respectively,
${\cal T}_{\mu \nu}(q) = g_{\mu \nu} - q_\mu q_\nu/q^2$,
${\cal L}_{\mu \nu}(q) = q_\mu q_\nu/q^2$\/.

The renormalized fermion propagator is as in Eq. (\ref{fermprop_formal}).
$\Gamma_\nu(k,p)$ is the renormalized proper vertex; $q = k - p$
is the photon momentum.
The renormalization point for $S$, $\Gamma$,
and $D$ is $p^2 = \mu^2$, however, we will not always write it
explicitly.

The ``naively'' regularized self-energy
(under a regularization scheme with parameter $\Lambda$) is
\begin{equation}
  \Sigma^\Lambda(p) =  i Z_1(\mu,\Lambda) e^2
		       \int^\Lambda \frac{d^4k}{(2 \pi)^4}
		        \gamma^\mu S(k) \Gamma^\nu(k,p)
			D_{\mu \nu}(q)
\label{Eq_naive}
\end{equation}
(where we write the regularization as though it were a momentum
cutoff, but it need not be).
If $\Gamma^\nu(k,p)$ satisfies the WTI,
$
  q_\nu \Gamma^\nu(k,p) =  S^{-1}(k) - S^{-1}(p)\:,
$
we can rewrite the DSE as
\begin{eqnarray}
  \Sigma^\Lambda(\mu,\Lambda;p)
    & = & i Z_1(\mu,\Lambda) e^2
	    \int^\Lambda \frac{d^4k}{(2 \pi)^4}
		     \gamma^\mu S(k) \Gamma^\nu(k,p)
		     D^{\rm T}_{\mu \nu}(q)
						            \nonumber\\
    & & \;\; - i Z_1(\mu,\Lambda) e^2
		 \int^\Lambda \frac{d^4k}{(2 \pi)^4}
		     \gamma^\mu S(k) \Gamma^\nu(k,p)
		     \xi \frac{q_\mu q_\nu}{q^2} \frac{1}{q^2}
    \\
    & = & i Z_1(\mu,\Lambda) e^2
	    \int^\Lambda \frac{d^4k}{(2 \pi)^4}
		     \gamma^\mu S(k) \Gamma^\nu(k,p)
		     D^{\rm T}_{\mu \nu}(q)     \nonumber\\
    &   & \;\; - \;\fbox{$ \displaystyle
	   i \xi Z_1(\mu,\Lambda) e^2 \int^\Lambda \frac{d^4k}{(2 \pi)^4}
		    \frac{\not\!q}{q^2} \frac{1}{q^2}
	                $}
	    \nonumber\\
    &   & + i \xi Z_1(\mu,\Lambda) e^2
	      \int^\Lambda \frac{d^4k}{(2 \pi)^4}
		     \frac{\not\!q}{q^2} S(k) S^{-1}(p)
		     \frac{1}{q^2}                       \nonumber\\
\label{Eq_box}
\end{eqnarray}

The boxed integral is odd in $q$, and should vanish in any
translationally invariant regularization scheme;
otherwise, it contributes and we expect it to destroy the
gauge-covariance of $\Sigma^\Lambda$.
However, we can define a ``gauge-improved'' self-energy by cancelling
this undesirable term
\begin{equation}
  \Sigma'(\mu,\Lambda; p) \equiv \Sigma^\Lambda(\mu,\Lambda; p)
	       + i \xi Z_1(\mu,\Lambda)e^2
		 \int^\Lambda \frac{d^4k}{(2 \pi)^4}
		   \frac{\not\!q}{q^2} \frac{1}{q^2}
\end{equation}
and since the added integral is Dirac-odd, upon decomposing into
scalar and spinor parts as in Eq. (\ref{decompose}) we have
\begin{eqnarray}
  \Sigma'_s(\mu,\Lambda; p^2) & = & \Sigma^\Lambda_s(\mu,\Lambda; p^2)
    \:, \\
  \Sigma'_d(\mu,\Lambda; p^2) & = & \Sigma^\Lambda_d(\mu,\Lambda; p^2)
    + i \xi Z_1(\mu,\Lambda) e^2 \int^\Lambda \frac{d^4k}{(2 \pi)^4}
            \frac{p \cdot q}{p^2 q^4} \;.
\end{eqnarray}
This modification combined with the C-P vertex is the {\it Ansatz} used in
all calculations in this work unless explicitly stated otherwise.

Converting to Euclidean metric
(but suppressing the ``Euclidean'' subscript on momenta for convenience)
gives
\begin{eqnarray}
  \Sigma'_d(\mu,\Lambda; p^2)
    & = & \Sigma^\Lambda_d(\mu,\Lambda; p^2)
          - \xi Z_1(\mu,\Lambda) e^2
	    \int^\Lambda \frac{d^4k}{(2 \pi)^4}
                \frac{p \cdot q}{p^2 q^4}\:.
\end{eqnarray}
Since in our case the regularization is an ultraviolet cutoff
in momentum, we have
\begin{equation}
  \Sigma'_d(\mu,\Lambda; p^2) = \Sigma^\Lambda_d(\mu,\Lambda; p^2)
              - \xi Z_1(\mu,\Lambda) \frac{\alpha}{2 \pi^2}
		\int^{\Lambda^2} k^2 dk^2
		\int_0^\pi \sin^2 \theta d \theta
		  \frac{p \cdot q}{p^2 q^4} \;.
\end{equation}
Introducing the variables
$x = p^2$, $y = k^2$, $z=(k-p)^2=x+y-2\sqrt{xy}\cos\theta$,
$x_> = \max(x,y)$, $x_< = \min(x,y)$, 
we have
\begin{eqnarray}
  \Sigma'_d(\mu,\Lambda; p^2)
    & = & \Sigma^\Lambda_d(\mu,\Lambda; p^2)
          - \xi Z_1(\mu,\Lambda) \frac{\alpha}{2 \pi^2}
              \int^{\Lambda^2} y dy
	      \int_0^\pi \sin^2 \theta d \theta
	        \frac{\sqrt{xy} \cos \theta - x}{x z^2}
		\nonumber\\
    & = & \Sigma^\Lambda_d(\mu,\Lambda; p^2) -
                   Z_1(\mu,\Lambda) \frac{\alpha \xi}{2 \pi^2}
		\:\frac{\pi}{2} \: \frac{1}{x}
		\:\int_0^{\Lambda^2} dy
		  \frac{(x_< - x) y}{(x_> - x_<) x_>}
                \nonumber\\
    & = & \Sigma^\Lambda_d(\mu,\Lambda; p^2) +
                 Z_1(\mu,\Lambda) \frac{\alpha \xi}{8 \pi}\:,
\end{eqnarray}
Thus the gauge covariance correction that we use here is to cancel the boxed
term in Eq.~(\ref{Eq_box}) by adding $Z_1(\mu,\Lambda)\alpha\xi/8\pi$
to the naive regularized self-energy in Eq.~(\ref{Eq_naive}) and this is
how it is implemented in our program.  The effects of including and not
including this gauge covariance correction can be studied numerically.


\end{appendix}


\newpage


\vspace{8.5cm}

\vspace{8.5cm}

\begin{table}
  \caption{ Renormalization constant $Z_2(\mu,\Lambda)$, bare masses
    $m_0(\Lambda)$, and mass renormalization $Z_m(\mu,\Lambda)$, as
    a function of UV cutoff for $\alpha = 1.15$ in the Landau gauge
    ($\xi = 0$).  All solutions are with renormalization point
    $\mu^2 = 1.00\times 10^{8}$ and renormalized mass $m(\mu) = 400.0$}
  \setdec 0.0000
  \begin{tabular}{lr@{.}lr@{.}lr@{.}l}
    \multicolumn{1}{c}{$\Lambda^2$}         &
    \multicolumn{2}{c}{$Z_2(\mu,\Lambda)$}  &
    \multicolumn{2}{c}{$m_0(\Lambda)$}      &
    \multicolumn{2}{c}{$Z_m(\mu,\Lambda)$}  \\
  \tableline
      $1\times 10^{8}$  & 0&9999135 &  2 & $306 \times 10^{2}$
          & 5 & $765 \times 10^{-1}$ \\
      $1\times 10^{9}$  & 0&9998483 &  5 & $358 \times 10^{1}$
          & 1 & $339 \times 10^{-1}$ \\
      $1\times 10^{10}$ & 0&9998468 &  4 & 443
          & 1&$111 \times 10^{-2}$ \\
      $1\times 10^{11}$ & 0&9998469 & -3 & 932
          & -9&$831 \times 10^{-3}$ \\
      $1\times 10^{12}$ & 0&9998469 & -2 & 847
          & -7 & $117 \times 10^{-3}$ \\
      $1\times 10^{13}$ & 0&9998469 & -1 & 182
          & -2 & $954 \times 10^{-3}$ \\
      $1\times 10^{14}$ & 0&9998469 & -3 & $408 \times 10^{-1}$
          & -8 & $520 \times 10^{-4}$ \\
      $1\times 10^{15}$ & 0&9998469 & -5 & $390 \times 10^{-2}$
          & -1 & $348 \times 10^{-4}$ \\
      $1\times 10^{16}$ & 0&9998469 &  1 & $043 \times 10^{-2}$
          &  2 & $607 \times 10^{-5}$ \\
      $1\times 10^{17}$ & 0&9998469 &  1 & $276 \times 10^{-2}$
          &  3 & $191 \times 10^{-5}$ \\
      $1\times 10^{18}$ & 0&9998469 &  6 & $171 \times 10^{-3}$
          &  1 & $543 \times 10^{-5}$ \\
      $1\times 10^{19}$ & 0&9998469 &  2 & $042 \times 10^{-3}$
          &  5 & $105 \times 10^{-6}$ \\
  \end{tabular}
  \label{Landau_tbl}
\end{table}

\begin{table}
  \caption{ Renormalization constant $Z_2(\mu,\Lambda)$, bare masses
    $m_0(\Lambda)$, and mass renormalization $Z_m(\mu,\Lambda)$, as
    a function of UV cutoff for $\alpha = 1.15$ in the gauge with
    $\xi = 0.25$.  All solutions are with renormalization point
    $\mu^2 = 1.00\times 10^{8}$ and renormalized mass $m(\mu) = 400.0$}
  \setdec 0.0000
  \begin{tabular}{lr@{.}lr@{.}lr@{.}l}
    \multicolumn{1}{c}{$\Lambda^2$}         &
    \multicolumn{2}{c}{$Z_2(\mu,\Lambda)$}  &
    \multicolumn{2}{c}{$m_0(\Lambda)$}      &
    \multicolumn{2}{c}{$Z_m(\mu,\Lambda)$}  \\
  \tableline
      $1 \times 10^{8}$  & 0&999943  &  2 & $239 \times 10^{2}$
	  &  5 & $598 \times 10^{-1}$   \\
      $1 \times 10^{9}$  & 0&9486    &  4 & $918 \times 10^{1}$
	  &  1 & $229 \times 10^{-1}$   \\
      $1 \times 10^{10}$ & 0&8999    &  2 & 034
	  &  5 & $085 \times 10^{-3}$   \\
      $1 \times 10^{11}$ & 0&8537    & -4 & 898
	  & -1 & $225 \times 10^{-2}$   \\
      $1 \times 10^{12}$ & 0&8099    & -3 & 102
	  & -7 & $756 \times 10^{-3}$   \\
      $1 \times 10^{13}$ & 0&7683    & -1 & 193
	  & -2 & $981 \times 10^{-3}$   \\
      $1 \times 10^{14}$ & 0&7289    & -3 & $059 \times 10^{-1}$
	  & -7 & $647 \times 10^{-4}$   \\
      $1 \times 10^{15}$ & 0&6915    & -2 & $886 \times 10^{-2}$
	  & -7 & $214 \times 10^{-5}$   \\
      $1 \times 10^{16}$ & 0&6560    &  2 & $145 \times 10^{-2}$
	  &  5 & $362 \times 10^{-5}$   \\
      $1 \times 10^{17}$ & 0&6224    &  1 & $609 \times 10^{-2}$
	  &  4 & $023 \times 10^{-5}$   \\
      $1 \times 10^{18}$ & 0&5904    &  6 & $679 \times 10^{-3}$
	  &  1 & $670 \times 10^{-5}$   \\
  \end{tabular}
  \label{gp025_tbl}
\end{table}

\begin{table}
  \caption{ Renormalization constant $Z_2(\mu,\Lambda)$, bare masses
    $m_0(\Lambda)$, and mass renormalization $Z_m(\mu,\Lambda)$, as
    a function of UV cutoff for $\alpha = 1.15$ in the gauge with
    $\xi = 0.5$.  All solutions are with renormalization point
    $\mu^2 = 1.00\times 10^{8}$ and renormalized mass $m(\mu) = 400.0$}
  \setdec 0.0000
  \begin{tabular}{lr@{.}lr@{.}lr@{.}l}
    \multicolumn{1}{c}{$\Lambda^2$}         &
    \multicolumn{2}{c}{$Z_2(\mu,\Lambda)$}  &
    \multicolumn{2}{c}{$m_0(\Lambda)$}      &
    \multicolumn{2}{c}{$Z_m(\mu,\Lambda)$}  \\
  \tableline
      $1 \times 10^{8}$  & 0&99997 &  2 & $176 \times 10^{2}$
	  &  5 & $441 \times 10^{-1}$ \\
      $1 \times 10^{9}$  & 0&8999  &  4 & $513 \times 10^{1}$
	  &  1 & $128 \times 10^{-1}$ \\
      $1 \times 10^{10}$ & 0&8099  & -1 & $455 \times 10^{-1}$
	  & -3 & $638 \times 10^{-4}$ \\
      $1 \times 10^{11}$ & 0&7289  & -5 & 736
	  & -1 & $434 \times 10^{-2}$ \\
      $1 \times 10^{12}$ & 0&6560  & -3 & 299
	  & -8 & $248 \times 10^{-3}$ \\
      $1 \times 10^{13}$ & 0&5904  & -1 & 181
	  & -2 & $954 \times 10^{-3}$ \\
      $1 \times 10^{14}$ & 0&5314  & -2 & $653 \times 10^{-1}$
	  & -6 & $632 \times 10^{-4}$ \\
      $1 \times 10^{15}$ & 0&4783  & -3 & $676 \times 10^{-3}$
	  & -9 & $190 \times 10^{-6}$ \\
      $1 \times 10^{16}$ & 0&4304  &  3 & $151 \times 10^{-2}$
	  & 7 & $877 \times 10^{-5}$ \\
      $1 \times 10^{17}$ & 0&3874  &  1 & $870 \times 10^{-2}$
	  & 4 & $674 \times 10^{-5}$ \\
  \end{tabular}
  \label{gp050_tbl}
\end{table}

\begin{table}
  \caption{Critical parameters for three choices of gauge,
    $\xi=0$, 0.25, and 0.50.
    These are extracted from nonlinear fits
    to the data in Fig.\ \protect\ref{crit_curves}\/,
    using the form in Eq.\ (\protect\ref{crit_functionalform}).
    }
  \setdec 0.0000
  \begin{tabular}{clll}
    Parameter & Landau ($\xi=0$) & $\xi = 0.25$ & $\xi = 0.5$ \\
    \tableline
      $c$        & $2.877 \pm .027$      & $2.858 \pm .043$      &
        $2.851 \pm  .055$ \\
      $\alpha_c$ & $0.93307 \pm  .00023$ & $0.92076 \pm .00048$  &
	$0.90946 \pm  .00071$ \\
      $\beta$    & $0.512 \pm  .003$     & $0.514 \pm  .005$     &
	$0.516 \pm  .007$ \\
      $M$        & $154.3 \pm 5.2$       & $148.5 \pm 7.7 $      &
	$145.4 \pm 9.4$ \\
    \tableline
      $\chi^2/{\rm DOF}$ &  0.0959       & .0388                 &
	.0211  \\
  \end{tabular}
  \label{crit_fits_tbl}
\end{table}


\begin{references}
\bibitem{TheReview}  C.D.~Roberts and A.G.~Williams,
    {\it Dyson-Schwinger Equations and their Application to Hadronic
    Physics\/}, in
    {\it Progress in Particle and Nuclear Physics, Vol.~33}
    (Pergamon Press, Oxford, 1994), p.~477.
\bibitem{MiranskReview} V.\ A.\ Miranskii,
    {\it Dynamical Symmetry Breaking in Quantum Field Theories},
    (World Scientific, Singapore, 1993).
\bibitem{FGMS} P.\ I.\ Fomin, V.\ P.\ Gusynin, V.\ A.\ Miransky and
    Yu.\ A.\ Sitenko, Riv. Nuovo Cim. {\bf 6}, 1 (1983).
\bibitem{qed4_hw} F.T.\ Hawes and A.G.\ Williams,
    Phys.\ Rev.\ D {\bf 51}, 3081 (1995).
\bibitem{Bare} R.~Haag and Th.A.~Maris, Phys.~Rev.~{\bf 132}, 2325
    (1963); Th.A.J.~Maris, V.E.~Herscovitz and D.~Jacob,
    Phys.~Rev.~Lett. {\bf 12}, 313 (1964); T.~Nonoyama and M.~Tanabashi,
    Prog.~Theor.~Phys.~{\bf 81}, 209 (1989).
\bibitem{Miransk1} P.\ I.\ Fomin, V.\ P.\ Gusynin, and V.\ A.\ Miransky,
    Phys.\ Lett.\ {\bf 78B}, 136 (1978).
\bibitem{Miransk2} V.\ A.\ Miransky, Sov. Phys. JETP {\bf 61} (5),
    905 (1985).
\bibitem{Miransk3} V.~A.~Miransky, Nuovo Cimento {\bf 90A}, 149 (1985).
\bibitem{Rakow} P.\ E.\ L.\ Rakow, Nucl. Phys. {\bf B356}, 27 (1991).
\bibitem{WTI} J.C.~Ward, Phys.~Rev.~{\bf 78}, 124 (1950);
    Y.~Takahashi, Nuovo Cimento {\bf 6}, 370 (1957).
\bibitem{ABKHN} K.-I.\ Aoki, M.\ Bando, T.\ Kugo, K.\ Hasebe
    and H.\ Nakatani, Prog.\ Theor.\ Phys.\ {\bf 81}, 866 (1989).
\bibitem{CPIV} D.C.~Curtis and M.R.~Pennington, Phys. Rev. D{\bf 48},
    4933 (1993).
\bibitem{KKM} K.-I.~Kondo, Y.~Kikukawa and H.~Mino,
    Phys.~Lett.~{\bf 220B}, 270 (1989).
\bibitem{bad_Gamma} D.~Atkinson, P.~W.~Johnson and K.~Stam,
    Phys.~Lett.~{\bf 201B}, 105 (1988).
\bibitem{BC} J.S.~Ball and T.W.~Chiu, Phys. Rev. D{\bf 22},
    2542 (1980); {\it ibid.\/}, 2550 (1980).
\bibitem{King} J.E.~King, Phys. Rev. D{\bf 27}, 1821 (1983).
\bibitem{gaugetech} P.~Rembiesa, Phys.~Rev.~D{\bf 41}, 2009 (1990).
\bibitem{HaeriQED} B.~Haeri, Phys. Rev. D{\bf 43}, 2701 (1991).
\bibitem{CPI} D.C.~Curtis and M.R.~Pennington, Phys. Rev. D{\bf 42},
    4165 (1990).
\bibitem{CPII} D.C.~Curtis and M.R.~Pennington, Phys. Rev. D{\bf 44},
    536 (1991).
\bibitem{CPIII} D.C.~Curtis and M.R.~Pennington, Phys. Rev. D{\bf 46},
    2663 (1992).
\bibitem{dongroberts} Z.~Dong, H.~Munczek, and C.D.~Roberts,
    Phys. Lett. {\bf 333B}, 536 (1994).
\bibitem{BP1} A.~Bashir and M.~R.~Pennington,
    Phys.\ Rev.\ D {\bf 50}, 7679 (1994).
\bibitem{BP2} A.~Bashir and M.~R.~Pennington,
    Phys.\ Rev.\ D {\bf 53}, 4694 (1996).
\bibitem{Craig_1} C.D.~Roberts, {\it Schwinger Dyson Equations:
    Dynamical chiral symmetry breaking and Confinement\/}, in
    {\it QCD Vacuum Structure\/}, edited by H.~M.~Fried and
    B.~M\"{u}ller (World Scientific, Singapore, 1993).
\bibitem{Kiz_et_al} A.\ Kizilers\"{u}, M.\ Reenders, and
    M.\ R.\ Pennington, Phys.\ Rev.\ D {\bf 52}, 1242 (1995).
\bibitem{ABGPR} D.~Atkinson, J.C.R.~Bloch, V.P.~Gusynin, M.~R.~
    Pennington, and M.~Reenders, Phys. Lett. {\bf 329B}, 117 (1994).
\bibitem{dim_reg} L.\ von Smekal, P.~A.\ Amundsen, and R.\ Alkofer,
    Nucl.\ Phys.\ {\bf A529}, 633 (1991);
    M.\ Becker, ``Nichtperturbative Strukturuntersuchungen der
      QED mittels gen\"{a}herter Schwinger-Dyson-Gleichungen
      in Dimensioneller Regularisierung,'', Ph.D.\ dissertation,
      W.~W.~U.\ M\"{u}nster,1995.
\bibitem{IZ} C.~Itzykson and J.B.~Zuber, {\it Quantum Field Theory\/},
    (McGraw-Hill, New York, 1980).
\bibitem{Rothe} H.\ J.\ Rothe,
    {\it Latticd Gauge Theories:  an Introduction\/},
    (World Scientific, Singapore, 1992).
\bibitem{Atk+Fry} D.\ Atkinson and M.P.\ Fry,
    Nucl.\ Phys.\ {B156}, 301 (1979).
\bibitem{LKTF} L.~D.~ Landau and I.\ M.\ Khalatnikov,
    Sov.\ Phys.\ JETP {\bf 2}, 69 (1956)
    [translation of Zhur. Eksptl. i Teoret. Fiz.\ {\bf 29}, 89 (1955)];
    K.\ Johnson and B.\ Zumino,
    Phys.\ Rev.\ Lett.\ {\bf 3}, 351 (1959).
\bibitem{Holdom} B.~Holdom, Phys.\ Rev.\ Lett.\ {\bf 62}, 997 (1989);
    (R) {\it ibid.}\/ {\bf 63}, 1889 (1989);
\bibitem{Mahanta1} U.~Mahanta, Phys.\ Lett.\ {\bf 225B}, 181 (1989).
\bibitem{Mahanta2} U.~Mahanta,
    Phys.\ Rev.\ Lett.\ {\bf 62}, 2349 (1989).
\bibitem{pirho_fr}  M.\ R.\ Frank and C.\ D.\ Roberts,
    Phys.\ Rev.\ D {\bf 53}, 390 (1996).
\end{references}
\end{document}